\documentclass[aps,pra,reprint,superscriptaddress]{revtex4-2}
\usepackage{amsmath,amssymb,amsfonts}
\usepackage{graphicx}
\usepackage{hyperref}
\usepackage{braket}
\usepackage{mathtools}
\usepackage{tikz}
\usepackage{bbold}
\usepackage{physics}
\usepackage{subfigure}
\usetikzlibrary{arrows.meta}
\hypersetup{
  colorlinks = true,
  linkcolor  = blue,
  citecolor  = blue,
  urlcolor   = blue,
}

\begin{document}

\title{Coherence-Controlled Quantum Zeno Dynamics from Exact Reset Maps}

\author{Jishad Kumar}
\affiliation{%
  MSP group, Department of Applied Physics, Aalto University,\\
  P.O. Box 15600, FI-00076 Aalto, Espoo, Finland
}

\author{Achilleas Lazarides}
\affiliation{%
  Interdisciplinary Centre for Mathematical Modelling and Department of Mathematical Sciences,\\
  Loughborough University, Loughborough, Leicestershire LE11 3TU, United Kingdom
}

\author{Tapio Ala-Nissila}
\affiliation{%
  MSP group, Department of Applied Physics, Aalto University,\\
  P.O. Box 15600, FI-00076 Aalto, Espoo, Finland
}
\affiliation{%
  Interdisciplinary Centre for Mathematical Modelling and Department of Mathematical Sciences,\\
  Loughborough University, Loughborough, Leicestershire LE11 3TU, United Kingdom
}

\date{\today}

\begin{abstract}
We develop an exact framework for quantum Zeno and anti-Zeno dynamics in a broad class of open systems, whose microscopic Hamiltonians are quadratic in bosonic or fermionic operators. 
We treat the environment through an exact stroboscopic resetting scheme acting at the level of the single-particle density matrix (SPDM). Within this framework, we consider two cases: a repeated-interaction (RI) protocol, in which the environment block is rethermalized and all system-environment coherences are erased after each step, and an evolving-correlation (EC) protocol, in which only the environment block is reset while system-environment coherences are preserved. For RI, we derive a general short-time Zeno law for the survival probability of a single-particle excitation and show that the corresponding decay rate scales linearly with the reset interval, implying Zeno freezing in the limit of infinitely frequent resets. Beyond the short-time regime, we formulate the RI dynamics directly in terms of the exact one-cycle propagator, which allows us to analyze finite-\(\tau\) anti-Zeno windows without additional approximations. For EC, we obtain a continuous-reset description in which the kept single-particle correlators obey a finite-dimensional linear differential equation. In this case the drift in the system block remains finite in the frequent-reset limit, so strict freezing is absent. We illustrate these results for a single fermionic level coupled to a semi-infinite tight-binding chain acting as a structured bath. Our results identify coherence erasure versus coherence retention as the key factor controlling the reset-induced Zeno physics.
\end{abstract}

\maketitle

\section{Introduction}

The famous quantum Zeno effect (QZE), the inhibition of quantum evolution by frequent measurements, was first analyzed in detail by Misra and Sudarshan in 1977~\cite{MisraSudarshan1977}. They showed that an unstable quantum state subject to sufficiently frequent projective observations can be ``frozen'' in its initial state (see also Refs.~\onlinecite{Schulman1998ContinuousAndPulsed,Fischer2001ObservationZenoAntiZeno,HomeWhitaker1997}
for early analyses of pulsed vs continuous monitoring and of the inverse
quantum Zeno effect), because the survival probability exhibits a universal quadratic short-time behavior. Subsequent work established that frequent observations can also \emph{accelerate} decay, leading to the so-called quantum anti-Zeno effect (AZE)~\cite{KofmanKurizki2000Nature,KofmanKurizki2000FortschrPhys,Ruseckas2004}. Numerous experiments have observed Zeno and anti-Zeno behavior in a variety of atomic, optical, and solid-state systems (see reviews in Refs.~\onlinecite{FacchiNakazatoPascazio2001,Venugopalan2012,Itano2009}
and experiments in Refs.~\onlinecite{Itano1990QuantumZeno,Fischer2001ObservationZenoAntiZeno,Streed2006ContinuousPulsedZeno}).

From a modern perspective, the QZE is not restricted to ideal projective measurements, but can arise from a wide class of dynamical interventions, including coupling to an ancillary system or environment, frequent dephasing, or dynamical decoupling sequences~\cite{FacchiPascazio2008JPA,FacchiPascazio2002PRL}. The common feature is that rapid interventions interrupt the coherent evolution and modify the effective spectral overlap between the system and its environment~\cite{KofmanKurizki2000Nature,Ruseckas2004}. In open quantum systems, this viewpoint naturally leads to the study of Zeno and anti-Zeno physics within the framework of system--bath models and master equations~\cite{ZhouJinFicek2017,Becker2021}.

In parallel, \emph{collision models} or \emph{repeated-interaction} schemes have emerged as a powerful microscopic representation of open-system dynamics~\cite{Ciccarello2022PhysRep,Cusumano2022Entropy}, and earlier works on collisional realizations of Markovian and non-Markovian dynamics~\cite{Giovannetti2012CollisionDM,Walter2016,Milz2018}. In these models the system interacts sequentially with ancillary units of the environment, each of which is reset to a reference state between collisions. Such models give rise to completely positive maps, Lindblad generators in suitable limits, and controlled deviations from Markovianity, and have been used to study quantum thermodynamics~\cite{Scarani2002PRL,Strasberg2017PRX,Strasberg2019PRL}, non-Markovian dynamics~\cite{Ciccarello2013PRA,Walter2016,Lorenzo2017OSID}, and quantum information processing~\cite{Giovannetti2005JPA,Burgarth2007PRA,Cattaneo2021PRL,Cakmak2019PRA}.

A particularly important role in resetting protocols in open quantum systems is played by exactly solvable models such as quadratic fermionic Hamiltonians~\cite{vieira2020}. In the noninteracting case, it is possible to study different resetting protocols at the single-particle density matrix (SPDM) level, enabling exact characterization of relaxation, transport, and dynamical fixed points without explicit modeling of microscopic reservoir dynamics. The reset protocols induce an effective interaction that leads to homogenized steady-states, which in some cases correspond to environmentally thermalized or pseudothermalized states. The fermionic model has also been generalized to the case of weak interactions, where new interaction-induced steady states emerge \cite{kumar2026interactionenabledhartreefixedpoints}. In another recent work \cite{Prositto_2025}, the authors studied a solvable model of a qubit interacting with a stream of thermalized ancilla qubits with a Heisenberg-type of Hamiltonian in the repeated-interaction protocol. Similar to the noninteracting case in Ref. \onlinecite{vieira2020}, they find steady states that are diagonal but not necessarily thermalized with the ancilla. 

Although both Zeno and anti-Zeno effects have been studied extensively in measurement-based, driven, and open-system settings, one fundamental question remains unanswered: \emph{what precisely determines the frequent-reset limit when the environment is repeatedly reinitialized?} This question is particularly relevant in collision models and repeated-interaction schemes, where resetting is not merely a mathematical device but a physically meaningful operation acting on environmental degrees of freedom. In such settings, different reset rules can look superficially similar (they may rethermalize the same environment block at the same rate) yet they need not produce the same steady states. We demonstrate here that a key role is played by how the system-environment coherences are treated in the reset protocol.

To this end, we extend the analysis of Ref. \onlinecite{vieira2020} to analyze Zeno dynamics in a general class of open quantum systems whose dynamics is generated by a quadratic Hamiltonian for the system and environment degrees of freedom. Rather than postulating a master equation or making any assumptions on Markovianity, we work directly at the SPDM level, where the dynamics is linear and exactly solvable. We assume a stroboscopic scheme in which the global system-environment state evolves unitarily for a fixed time $\tau$ and is then subjected to a deterministic ``reset'' operation on the environment degrees of freedom.
We follow Ref. \onlinecite{vieira2020} and compare two physically motivated reset protocols (cf. Fig. 1 in Ref. \onlinecite{vieira2020}): (i) A \emph{repeated-interaction} (RI) protocol, in which the environment is reinitialized to a reference thermal state and all system-environment coherences are erased after each step. (ii) An \emph{evolving-correlation} (EC) protocol, in which the environment is reset as in RI, but system-environment coherences are retained and evolve unitarily between resets (Sec.~\ref{sec:model}). Both protocols are natural in the language of collision models~\cite{Ciccarello2022PhysRep,Cusumano2022Entropy}, but they lead to very different Zeno phenomenology.

For the RI protocol, we derive a general short-time Zeno law for the survival probability of a single-particle excitation in the system. We show that the per-cycle change in the survival probability is of order $\tau^2$, with an explicit prefactor expressed in terms of the microscopic couplings. This implies an effective decay rate $\Gamma_{\rm RI}(\tau)\propto \tau$ and therefore strict Zeno freezing as $\tau\to 0$ (Sec.~\ref{sec:RIshorttime}). Beyond the universal short-time regime, we formulate the RI dynamics directly in terms of the exact one-cycle propagator. For the single-level problem, the effective decay rate is obtained from the exact survival amplitude $U_{00}(\tau)$, which allows us to analyze both the strict Zeno regime at small $\tau$ and finite-$\tau$ anti-Zeno behavior within the same exact stroboscopic framework (Sec.~\ref{sec:RIshorttime}). For the EC protocol, we obtain a continuous-reset limit in which the vector of kept SPDM entries obeys a finite-dimensional linear differential equation. We show that the drift in the system block remains finite as $\tau\to 0$, implying the absence of strict Zeno freezing, although Zeno-like slow-down and anti-Zeno acceleration can still occur at finite $\tau$ (Sec.~\ref{sec:EC}). The essential distinction between the two protocols is that RI erases system-environment coherences after each reset step, whereas EC retains them and allows them to feed back into the system dynamics. \emph{Thus the presence or absence of coherence erasure is the key control parameter that selects the asymptotic Zeno behavior, and the anti-Zeno effect is a direct consequence of the exact stroboscopic map rather than of an auxiliary approximation scheme.}

We illustrate these general results on a concrete example: a single fermionic level coupled to a semi-infinite tight-binding chain. For RI we compute the exact effective decay rate as a function of the reset interval $\tau$ and demonstrate both strict Zeno suppression and finite-$\tau$ anti-Zeno windows. For EC we compare the decay rates inferred from the continuous generator and from the exact stroboscopic map. We therefore arrive at the following: erasing versus preserving system-environment coherences is the organizing principle that determines the reset-induced Zeno physics. This provides a common exact framework for quantum Zeno dynamics, anti-Zeno enhancement, and collision-model resetting in quadratic open quantum systems.

\section{Quadratic model and stroboscopic resetting}
\label{sec:model}

\subsection{Quadratic Hamiltonian and single-particle density matrix}

We consider a set of fermionic or bosonic annihilation operators $\{\hat a_\alpha\}_{\alpha=1}^N$ satisfying the respective canonical anticommutation and commutation relations. The dynamics is generated by a quadratic Hamiltonian
\begin{equation}
    \hat H = \sum_{\alpha,\beta=1}^N \hat a^\dagger_\alpha\, M_{\alpha\beta}\, \hat a_\beta,
    \label{eq:Hquadratic}
\end{equation}
where $M$ is an $N\times N$ Hermitian matrix. Throughout we set $\hbar=1$. The single-particle density matrix
\begin{equation}
    \rho_{\alpha\beta}(t) \equiv \langle \hat a_\alpha^\dagger \hat a_\beta \rangle_t
    \label{eq:rho_def}
\end{equation}
fully characterizes Gaussian states and evolves under the unitary propagator
\begin{equation}
    U(t) = e^{-iMt}.
\end{equation}
Explicitly,
\begin{align}
    \rho(t) &= U^\dagger(t)\,\rho(0)\,U(t);\\
    \rho_{\alpha\beta}(t)
    &= \sum_{\alpha',\beta'} U^*_{\alpha\alpha'}(t)\,U_{\beta\beta'}(t)\,\rho_{\alpha'\beta'}(0).
    \label{eq:unitary_SPDM}
\end{align}
We partition the single-particle Hilbert space into a system sector $S$ and an environment sector $E$, such that
\begin{equation}
    \rho = \begin{pmatrix} \rho_{SS} & \rho_{SE}\\ \rho_{ES} & \rho_{EE}\end{pmatrix};
    \quad
    M = \begin{pmatrix} M_{SS} & M_{SE}\\ M_{ES} & M_{EE}\end{pmatrix}.
\end{equation}

\subsection{Stroboscopic evolution and reset map}
\label{subsec:stroboscopic_map}

We assume a stroboscopic protocol with step duration $\tau$: starting from time $t_n=n\tau$, the global state evolves unitarily to $t_{n+1}^- = t_n+\tau$ according to Eq.~\eqref{eq:unitary_SPDM}, and is then subjected to a deterministic reset operation acting on $E$ (and possibly on $SE$ coherences). The minus sign in $t^-$ denotes the final unitary evolution time before instantaneous reset. At the SPDM level, the reset is a linear map that overwrites a specified set of entries with prescribed values.

Let $R\subset \{1,\dots,N\}^2$ be the set of ordered index pairs $(\alpha,\beta)$ for which we reset $\rho_{\alpha\beta}$ to some fixed value $\rho^{(0)}_{\alpha\beta}$ after each unitary step. Let ${\cal K}=R^{\rm c}$ be the complementary set of index pairs that are \emph{kept} from step to step. We define a vector of kept SPDM entries
\begin{equation}
    V_i[n] \equiv \rho_{\alpha_i\beta_i}(t_n),
\end{equation}
where $i\mapsto(\alpha_i,\beta_i)\in{\cal K}$ is a fixed enumeration of the kept indices.

From Eq.~\eqref{eq:unitary_SPDM}, the unitary evolution over a single step gives \cite{vieira2020}
\begin{equation}
    \rho_{\alpha\beta}(t_{n+1}^-)
    = \sum_{\alpha',\beta'} U^*_{\alpha\alpha'}(\tau)\,U_{\beta\beta'}(\tau)\,\rho_{\alpha'\beta'}(t_n).
\end{equation}
We then apply the reset operation: for $(\alpha,\beta)\in R$, we set
\begin{equation}
    \rho_{\alpha\beta}(t_{n+1}) = \rho_{\alpha\beta}^{(0)},
\end{equation}
while for $(\alpha,\beta)\in{\cal K}$, we keep the unitary-updated value,
\begin{equation}
    \rho_{\alpha\beta}(t_{n+1}) = \rho_{\alpha\beta}(t_{n+1}^-).
\end{equation}
Thus for each kept pair $(\alpha_i,\beta_i)\in{\cal K}$,
\begin{align}
    V_i[n+1]
    &= \rho_{\alpha_i\beta_i}(t_{n+1}) \nonumber\\
    &= \sum_{(\alpha',\beta')\in{\cal K}} U^*_{\alpha_i\alpha'}(\tau)\,U_{\beta_i\beta'}(\tau)\,
    \rho_{\alpha'\beta'}(t_n)\nonumber\\
    &\quad+ \sum_{(\alpha',\beta')\in R} U^*_{\alpha_i\alpha'}(\tau)\,U_{\beta_i\beta'}(\tau)\,
    \rho_{\alpha'\beta'}^{(0)}.
\end{align}
We now define
\begin{align}
    D_{ij}(\tau) &\equiv U^*_{\alpha_i\alpha_j}(\tau)\,U_{\beta_i\beta_j}(\tau);\\
    C_i(\tau) &\equiv \sum_{(\alpha',\beta')\in R}
    U^*_{\alpha_i\alpha'}(\tau)\,U_{\beta_i\beta'}(\tau)\,\rho_{\alpha'\beta'}^{(0)}.
\end{align}
This yields the exact \emph{affine} stroboscopic map
\begin{equation}
    V[n+1] = D(\tau)\,V[n] + C(\tau).
    \label{eq:stroboscopic_map}
\end{equation}
Equation~\eqref{eq:stroboscopic_map} is valid for arbitrary quadratic Hamiltonian $M$, arbitrary Gaussian reset state $\rho^{(0)}$, and arbitrary choice of the reset set $R$.

\begin{figure}[t]
\centering
\resizebox{\columnwidth}{!}{%
\begin{tikzpicture}[>=Stealth,thick,font=\small]

\tikzset{
  sys/.style={draw=blue!60!black,
              rounded corners=2pt,
              minimum width=1.6cm,
              minimum height=0.9cm,
              fill=blue!10},
  env/.style={draw=orange!70!black,
              rounded corners=2pt,
              minimum width=1.6cm,
              minimum height=0.9cm,
              fill=orange!15}
}

\node[anchor=west,font=\normalsize] at (-1.8,3.0) {(a) RI};

\node[sys] (S0) at (0,2.4) {S};
\node[env] (E0) at (0,0.9) {E};
\node[font=\footnotesize] at (0,0.2) {before step};

\node[sys] (S1) at (3.2,2.4) {S};
\node[env] (E1) at (3.2,0.9) {E};
\node[font=\footnotesize] at (3.2,0.2) {after unitary $U(\tau)$};

\node[sys] (S2) at (6.4,2.4) {S};
\node[env] (E2) at (6.4,0.9) {E};
\node[font=\footnotesize] at (6.4,0.2) {after reset (RI)};

\draw[->] (S0.east) .. controls (1.0,2.4) .. (S1.west);
\draw[->] (E0.east) .. controls (1.0,0.9) .. (E1.west);
\node[font=\footnotesize] at (1.6,2.9) {$U(\tau)$};

\draw[->] (S1.east) .. controls (4.2,2.4) .. (S2.west);
\draw[->] (E1.east) .. controls (4.2,0.9) .. (E2.west);
\node[font=\footnotesize] at (4.8,2.9) {reset};

\draw[<->] (S1.south) -- (E1.north);
\node[font=\footnotesize,fill=white,inner sep=1pt] at (3.2,1.55)
  {$\rho_{SE},\rho_{ES}$};

\draw[<->,dashed] (S2.south) -- (E2.north);
\node[font=\footnotesize,fill=white,inner sep=1pt] at (6.4,1.55)
  {$\rho_{SE},\rho_{ES}=0$};

\node[anchor=west,font=\normalsize] at (-1.8,-0.4) {(b) EC};

\node[sys] (S3) at (0,-1.4) {S};
\node[env] (E3) at (0,-2.9) {E};
\node[font=\footnotesize] at (0,-3.6) {before step};

\node[sys] (S4) at (3.2,-1.4) {S};
\node[env] (E4) at (3.2,-2.9) {E};
\node[font=\footnotesize] at (3.2,-3.6) {after unitary $U(\tau)$};

\node[sys] (S5) at (6.4,-1.4) {S};
\node[env] (E5) at (6.4,-2.9) {E};
\node[font=\footnotesize] at (6.4,-3.6) {after reset (EC)};

\draw[->] (S3.east) .. controls (1.0,-1.4) .. (S4.west);
\draw[->] (E3.east) .. controls (1.0,-2.9) .. (E4.west);
\node[font=\footnotesize] at (1.6,-0.85) {$U(\tau)$};

\draw[->] (S4.east) .. controls (4.2,-1.4) .. (S5.west);
\draw[->] (E4.east) .. controls (4.2,-2.9) .. (E5.west);
\node[font=\footnotesize] at (4.8,-0.85) {reset};

\draw[<->] (S4.south) -- (E4.north);
\node[font=\footnotesize,fill=white,inner sep=1pt] at (3.2,-2.15)
  {$\rho_{SE},\rho_{ES}$};

\draw[<->] (S5.south) -- (E5.north);
\node[font=\footnotesize,fill=white,inner sep=1pt] at (6.4,-2.15)
  {$\rho_{SE},\rho_{ES}$ kept};

\node[font=\footnotesize,anchor=west] at (-0.4,-4.1)
  {Both protocols: $\rho_{EE} \to \rho_{EE}^{(0)}$ after each step.};

\end{tikzpicture}
}
\caption{Schematic comparison between the repeated-interaction (RI) and evolving-correlation (EC) protocols. In each stroboscopic step the global quadratic system $S+E$ evolves unitarily under $U(\tau)$, followed by a reset of the environment block $\rho_{EE}$ to a fixed reference state $\rho_{EE}^{(0)}$. In RI (panel a), system-environment coherences $\rho_{SE}$ and $\rho_{ES}$ are erased after each reset. In EC (panel b), $\rho_{SE}$ and $\rho_{ES}$ are retained and evolve continuously during the dynamics.}
\label{fig:protocols}
\end{figure}
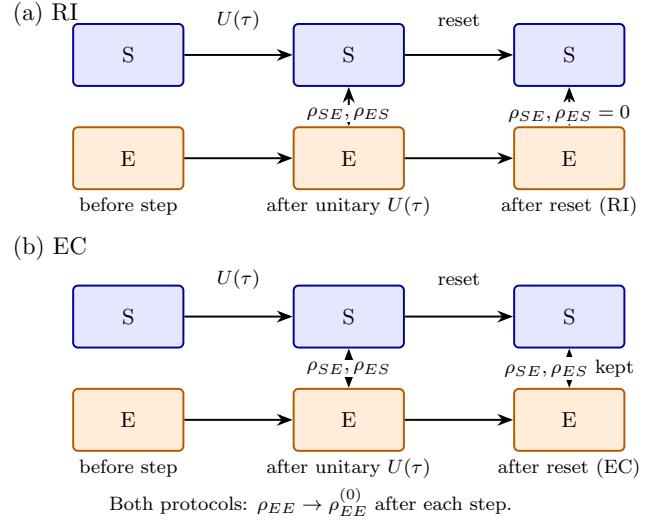

\subsection{RI and EC reset protocols at SPDM level}

We take $\rho(t_n)$ to be the SPDM
\emph{immediately after} the $n$th reset (i.e., $\rho(t_n)\equiv\rho(t_n^+)$), and $\rho(t_{n+1}^-)$ to be the SPDM \emph{just before} the next reset, obtained by unitary evolution over one stroboscopic step of duration $\tau$,
\begin{equation}
  \rho(t_{n+1}^-)
  = U(\tau)\,\rho(t_n)\,U^\dagger(\tau),
  \label{eq:unitary_step_block}
\end{equation}
which is equivalent to Eq.~\eqref{eq:unitary_SPDM} written in block form. The reset operation is then a linear map acting on the entries of $\rho(t_{n+1}^-)$, producing the post-reset SPDM $\rho(t_{n+1})\equiv\rho(t_{n+1}^+)$.

It is convenient to write $\rho(t)$ in block form with respect to the system ($S$) and environment ($E$) subspaces,
\begin{equation}
  \rho(t)
  =
  \begin{pmatrix}
    \rho_{SS}(t) & \rho_{SE}(t)\\[2pt]
    \rho_{ES}(t) & \rho_{EE}(t)
  \end{pmatrix},
\end{equation}
where, for example, $\rho_{SS}$ collects matrix elements $\rho_{\alpha\beta}$ with $\alpha,\beta\in S$, and similarly for the
other blocks. The reset rules discussed below are special choices of the reset set $R$ and reset values $\rho^{(0)}_{\alpha\beta}$ introduced in Sec.~\ref{subsec:stroboscopic_map}: for each $(\alpha,\beta)\in R$, we overwrite
$\rho_{\alpha\beta}(t_{n+1})=\rho^{(0)}_{\alpha\beta}$ after the unitary step, while for $(\alpha,\beta)\in\mathcal K=R^{\rm c}$ we keep the unitary-updated value.

\textit{Repeated interactions (RI).} In the RI protocol the environment block and all system-environment coherences are overwritten at each reset, while the system block is left unchanged. In block notation, the reset from
$\rho(t_{n+1}^-)$ to $\rho(t_{n+1})$ is
\begin{align}
  &\rho_{SS}(t_{n+1}) = \rho_{SS}(t_{n+1}^-); \label{eq:RI_SS}\\
  &\rho_{SE}(t_{n+1}) = 0; \qquad
   \rho_{ES}(t_{n+1}) = 0; \label{eq:RI_SEES}\\
  &\rho_{EE}(t_{n+1}) = \rho_{EE}^{(0)}, \label{eq:RI_EE}
\end{align}
where $\rho_{EE}^{(0)}$ is the fixed SPDM of a reference environment state (e.g., a thermal Fermi sea), and the zeros in
Eq.~\eqref{eq:RI_SEES} correspond to choosing $\rho^{(0)}_{\alpha\beta}=0$ for all $(\alpha,\beta)$ with one index in $S$ and the other in $E$. In terms of the general reset notation, this amounts to taking $R$ to contain all $EE$, $SE$, and $ES$ index pairs, with $\rho^{(0)}_{EE}=\rho_{EE}^{(0)}$ and $\rho^{(0)}_{SE}=\rho^{(0)}_{ES}=0$. Physically, RI realizes a collision model in which the system stroboscopically interacts with fresh, uncorrelated environment copies every interval $\tau$.

\textit{Evolving correlations (EC).} In the EC protocol the environment block is still driven back to the reference state at each reset, but the system-environment coherences are \emph{kept} at their terminal unitary values rather than erased. The corresponding block-level reset rules are
\begin{align}
  &\rho_{SS}(t_{n+1}) = \rho_{SS}(t_{n+1}^-); \label{eq:EC_SS}\\
  &\rho_{SE}(t_{n+1}) = \rho_{SE}(t_{n+1}^-);\,\,\,
   \rho_{ES}(t_{n+1}) = \rho_{ES}(t_{n+1}^-); \label{eq:EC_SEES}\\
  &\rho_{EE}(t_{n+1}) = \rho_{EE}^{(0)}, \label{eq:EC_EE}
\end{align}
so that only the $EE$ block is overwritten, with the same reference matrix $\rho_{EE}^{(0)}$ as in RI, while $SS$, $SE$, and $ES$ are all in the kept set $\mathcal K$. In the language of Sec.~\ref{subsec:stroboscopic_map}, this corresponds to taking $R$ to contain only the $EE$ index pairs, with $\rho^{(0)}_{EE}=\rho_{EE}^{(0)}$ and all other entries left untouched by the reset. One may view EC as describing an environment that is itself weakly coupled to a super-environment which repeatedly ``cools'' its
single-particle occupancies without destroying system-environment correlations.

In both RI and EC, the stroboscopic dynamics of the kept SPDM entries is fully captured by the affine map~\eqref{eq:stroboscopic_map} \cite{vieira2020},
\begin{equation}
  V[n+1] = D(\tau)\,V[n] + C(\tau),
\end{equation}
with the specific choice of the reset set $R$ and reset values $\rho_{\alpha\beta}^{(0)}$ distinguishing the two protocols.

\section{Short-time Zeno dynamics in the RI protocol}
\label{sec:RIshorttime}
In this section we analyze the short-time behavior of the RI protocol and derive a general Zeno law for the survival probability of a single-particle excitation in the system. The key result is that after each step of duration $\tau$ the change in the relevant system population is of order $\tau^2$, which yields an effective decay rate proportional to $\tau$.

\subsection{Short-time expansion of the SPDM}

We first expand the unitary propagator $U(\tau)$ to second order in $\tau$:
\begin{align}
    U(\tau) &= I - iM\tau - \tfrac{1}{2}M^2\tau^2 + O(\tau^3);\label{eq:U_expansion}\\
    U^\dagger(\tau) &= I + iM\tau - \tfrac{1}{2}M^2\tau^2 + O(\tau^3).
\end{align}
Inserting into $\rho' = U^\dagger(\tau)\rho\,U(\tau)$, we obtain
\begin{align}
    \rho'
    &= \Big(I + iM\tau - \tfrac{1}{2}M^2\tau^2\Big)\,\rho\,
       \Big(I - iM\tau - \tfrac{1}{2}M^2\tau^2\Big) \nonumber\\
    &\quad+ O(\tau^3)\nonumber\\
    &= \rho + iM\rho\tau - i\rho M\tau - \tfrac{1}{2}M^2\rho\tau^2
       - \tfrac{1}{2}\rho M^2\tau^2 \nonumber\\
    &\quad+ M\rho M\tau^2 + O(\tau^3).\label{eq:rho_prime_expanded}
\end{align}
The commutator and double commutator are
\begin{align}
    [M,\rho] &= M\rho - \rho M;\\
    [M,[M,\rho]] &= M^2\rho + \rho M^2 - 2M\rho M.
\end{align}
Comparing with Eq.~\eqref{eq:rho_prime_expanded}, we recognize the standard Baker-Campbell-Hausdorff expansion
\begin{equation}
    \rho' = \rho + i[M,\rho]\tau - \tfrac{1}{2}[M,[M,\rho]]\tau^2 + O(\tau^3).
    \label{eq:rho_prime_commutator}
\end{equation}

\subsection{Single-level survival probability}

We now focus on the survival probability of a single-particle excitation in a particular system level, which we label by $0\in S$. We consider the case where the system sector consists of this single level; the environment can be arbitrary and may contain an arbitrary number of modes. Let
\begin{equation}
    P \equiv \rho_{00} = \langle \hat a_0^\dagger \hat a_0\rangle
\end{equation}
denote the occupation of level $0$ at the beginning of a stroboscopic cycle. For clarity we take the environment reset state to be diagonal in the eigenbasis of $M_{EE}$,
\begin{equation}
    \rho_{ee'}^{(0)} = n_e\delta_{ee'};\qquad e,e'\in E,
\end{equation}
with occupations $n_e$ determined by temperature and statistics. We also assume that the SPDM is diagonal at the beginning of the cycle,
\begin{equation}
    \rho_{\alpha\beta}(t_n) = \rho_\alpha\,\delta_{\alpha\beta},
\end{equation}
with $\rho_0=P$ and $\rho_e=n_e$ for $e\in E$; system-environment coherences vanish at $t_n$.

Under unitary evolution to $t_{n+1}^-$, the occupation of level $0$ changes according to Eq.~\eqref{eq:rho_prime_commutator},
\begin{equation}
    \rho'_{00} = \rho_{00} + i[M,\rho]_{00}\tau - \tfrac{1}{2}[M,[M,\rho]]_{00}\tau^2 + O(\tau^3).
\end{equation}
For a diagonal SPDM, one has
\begin{equation}
    [M,\rho]_{\alpha\beta} = (\rho_\beta - \rho_\alpha)M_{\alpha\beta},
\end{equation}
so that in particular $[M,\rho]_{00}=0$ and the linear term vanishes identically. The leading short-time change of $P$ is therefore controlled by the double commutator.

In Appendix~\ref{app:double_commutator} we show explicitly that for a diagonal SPDM,
\begin{equation}
    [M,[M,\rho]]_{00} = 2\sum_{\gamma} |M_{0\gamma}|^2(\rho_{00}-\rho_{\gamma\gamma}),
    \label{eq:doble_commutator1}
\end{equation}
where the sum runs over all single-particle indices $\gamma$. Since the term with $\gamma=0$ vanishes identically ($\rho_{00}-\rho_{00}=0$), this can be written as
\begin{equation}
    [M,[M,\rho]]_{00} = 2\sum_{x\in E}|M_{0x}|^2\big(\rho_{00}-n_x\big),
    \label{eq:double_comm_identity}
\end{equation}
with the sum restricted to environment indices $x\in E$.

In the RI protocol, the reset step does not modify the system block: by Eq.~\eqref{eq:RI_SS} we have
$\rho_{00}(t_{n+1})=\rho'_{00}$. Thus the change in $P$ over one full cycle is
\begin{align}
    \Delta P &\equiv \rho_{00}(t_{n+1}) - \rho_{00}(t_n)\nonumber\\
    &= -\frac{\tau^2}{2}\,[M,[M,\rho]]_{00} + O(\tau^3)\nonumber\\
    &= -\tau^2 \sum_{x\in E} |M_{0x}|^2\big(P - n_x\big) + O(\tau^3).
    \label{eq:deltaP_RI}
\end{align}

For definiteness, consider the decay of a single fermionic excitation in level $0$ into an empty bath, $n_x=0$ for all $x\in E$. Then each cycle updates the survival probability as
\begin{equation}
    P_{n+1} = P_n - A\,\tau^2 P_n + O(\tau^3),
    \qquad A \equiv \sum_{x\in E} |M_{0x}|^2.
\end{equation}
Neglecting $O(\tau^3)$ corrections, this yields
\begin{equation}
    P_n \simeq (1-A\tau^2)^n.
\end{equation}
Letting $t=n\tau$ and taking the continuum limit with fixed $t$ and $\tau\to 0$, we find
\begin{equation}
    P(t) \simeq \big(1-A\tau^2\big)^{t/\tau}
    = e^{t\ln(1-A\tau^2)/\tau}
    \simeq e^{-A t\tau},
\end{equation}
where we used $\ln(1-x)=-x+O(x^2)$. Thus the effective decay rate is
\begin{equation}
    \Gamma_{\rm RI}(\tau) = A\,\tau + O(\tau^2).
    \label{eq:Gamma_RI_small_tau}
\end{equation}
In particular, $\Gamma_{\rm RI}(\tau)\to 0$ linearly as $\tau\to 0$, and the survival probability $P(t)$ tends to unity for any fixed finite $t$ in this limit: the excitation is strictly frozen in the ideal Zeno limit.

For a finite-temperature fermionic bath with occupations $n_x$, Eq.~\eqref{eq:deltaP_RI} implies that, to leading order in $\tau$, the effective decay rate out of the initially occupied level is
\begin{equation}
    \Gamma_{\rm RI}(\tau) = \tau\sum_{x\in E}|M_{0x}|^2\big(1-n_x\big) + O(\tau^2),
    \label{eq:Gamma_RI_fermion}
\end{equation}
where $1-n_x$ implements Pauli blocking of transitions into occupied bath modes. For bosons, the occupation factors are replaced by $1+n_x$.

\subsection{Single-level reduction of the RI map}

We focus on the case where the system sector $S$ consists of a single level labeled $0\in S$,
while the environment sector $E$ contains an arbitrary set of modes.
As in Sec.~II, the dynamics is generated by the quadratic Hamiltonian (1) with single-particle
matrix $M$ and propagator $U(t)=e^{-iMt}$, and the SPDM evolves according to Eq.~(\ref{eq:unitary_SPDM}). We will use the component form (\ref{eq:unitary_SPDM}) throughout to avoid any ambiguity in matrix ordering.

For later convenience we also introduce an eigenbasis of the environment block $M_{EE}$,
\begin{equation}
M_{EE}\,|k\rangle=\omega_k\,|k\rangle,\qquad k\in E,
\label{eq:env_eigen}
\end{equation}
and denote the system-environment couplings by
\begin{equation}
g_k \equiv M_{0k}\qquad (k\in E).
\label{eq:coupling_env}
\end{equation}
No weak-coupling assumption is made in what follows; Eqs.~(\ref{eq:env_eigen})-(\ref{eq:coupling_env}) are only notational.

Let
\begin{equation}
P_n \equiv \rho_{00}(t_n)=\langle \hat a_0^\dagger \hat a_0\rangle_{t_n}
\label{eq:old-returned}
\end{equation}
be the system occupation at the beginning of the $n$th cycle (immediately after the $n$th reset).
Under the unitary step $t_n\to t^-_{n+1}=t_n+\tau$, Eq.~(\ref{eq:unitary_SPDM}) gives for the $(0,0)$ SPDM element
\begin{equation}
\rho_{00}(t^-_{n+1})
=
\sum_{\alpha',\beta'\in S\cup E}
U^{*}_{0\alpha'}(\tau)\,U_{0\beta'}(\tau)\,\rho_{\alpha'\beta'}(t_n).
\tag{45}
\end{equation}
Now impose the \emph{RI reset rules} (\ref{eq:RI_SS})-(\ref{eq:RI_EE}): after each reset,
(i) the environment block is overwritten to the fixed Gaussian reference state $\rho^{(0)}_{EE}$,
(ii) system-environment coherences are erased, and
(iii) the system block is kept unchanged. 
In the present single-level setting this means that at time $t_n$ the only nonzero entries involving
the system index $0$ are $\rho_{00}(t_n)=P_n$, while $\rho_{0e}(t_n)=\rho_{e0}(t_n)=0$ for all $e\in E$.
Thus the SPDM at $t_n$ has the block form
\begin{align}
\rho_{\alpha'\beta'}(t_n)
&=
P_n\,\delta_{\alpha'0}\delta_{\beta'0}
+
\big(\rho^{(0)}_{EE}\big)_{\alpha'\beta'}\,\mathbf{1}_{\alpha'\in E}\mathbf{1}_{\beta'\in E};
\nonumber\\
\rho_{0e}(t_n)&=\rho_{e0}(t_n)=0.
\label{eq:components}
\end{align}
Substituting Eq.~(\ref{eq:components}) into Eq.~(\ref{eq:old-returned}) and separating the $\alpha'=\beta'=0$ contribution from the
purely-environment contributions yields the exact \emph{affine} recurrence
\begin{align}
P_{n+1}&=P(t_{n+1})=\rho_{00}(t_{n+1})\notag\\ &=\rho_{00}(t^-_{n+1})
=
a(\tau)\,P_n+b(\tau),
\label{eq:exact_affine_recurrence}
\end{align}
with coefficients
\begin{align}
a(\tau)&\equiv |U_{00}(\tau)|^2;
\\
b(\tau)&\equiv
\sum_{e,e'\in E}U^{*}_{0e}(\tau)\,U_{0e'}(\tau)\,\big(\rho^{(0)}_{EE}\big)_{ee'}.
\end{align}
Equation~(\ref{eq:exact_affine_recurrence}) is \emph{exact} for any finite-dimensional truncation (and remains exact in the infinite-bath limit provided $U(\tau)$ is defined as the corresponding one-particle propagator).

The solution of the affine map (\ref{eq:exact_affine_recurrence})  is elementary. Define the fixed point
$P_\ast(\tau)$ by $P_\ast=aP_\ast+b$, i.e.
\begin{align}
P_\ast(\tau) &=\frac{b(\tau)}{1-a(\tau)}\qquad (a(\tau)\neq 1);
\\
P_n &=P_\ast(\tau)+a(\tau)^n\big(P_0-P_\ast(\tau)\big).
\label{eq:solution_affine}
\end{align}
In particular, for an initially empty fermionic bath (or more generally any reset state with
$\rho^{(0)}_{EE}=0$) one has $b(\tau)=0$, so $P_n=a(\tau)^nP_0$ exactly.

\subsection{Exact decay rate from the map}

The stroboscopic relaxation of $P_n$ is governed by the single eigenvalue $a(\tau)=|U_{00}(\tau)|^2$.
It is therefore natural to define the \emph{exact RI decay rate extracted from the map} by
\begin{equation}
\Gamma_{\mathrm{eff}}(\tau)\equiv -\frac{1}{\tau}\ln a(\tau)
= -\frac{1}{\tau}\ln\!\Big(|U_{00}(\tau)|^2\Big).
\label{eq:gamma_decay_map}
\end{equation}
For $b(\tau)=0$ this implies the exact exponential form at stroboscopic times $t=n\tau$:
$P(t)=P_0\,e^{-\Gamma_{\mathrm{eff}}(\tau)t}$.

To verify consistency with the universal Zeno law in Sec.~\ref{sec:RIshorttime}, expand $U(\tau)=e^{-iM\tau}$ for small $\tau$. Using $U_{00}(\tau)=1-iM_{00}\tau-\frac{1}{2}(M^2)_{00}\tau^2+O(\tau^3)$ and
$(M^2)_{00}=M_{00}^2+\sum_{x\in E}|M_{0x}|^2$ (Hermiticity of $M$), one finds
\begin{equation}
a(\tau)=|U_{00}(\tau)|^2
=
1-\tau^2\sum_{x\in E}|M_{0x}|^2+O(\tau^3).
\label{eq:coeff_map_gamma}
\end{equation}
Inserting Eq.~(\ref{eq:coeff_map_gamma}) into Eq.~(\ref{eq:gamma_decay_map}) and using $\ln(1-\epsilon)=-\epsilon+O(\epsilon^2)$ gives
\begin{equation}
\Gamma_{\mathrm{eff}}(\tau)
=
\tau\sum_{x\in E}|M_{0x}|^2+O(\tau^2),
\label{eq:Gama_effective_affine}
\end{equation}
which reproduces the linear-in-$\tau$ Zeno scaling found in Eq.~(\ref{eq:Gamma_RI_small_tau}) for an empty fermionic bath. For finite-temperature fermions with reset occupations $n_x$, the \emph{initial short-time decay out of an
initially occupied level} acquires the Pauli factor $(1-n_x)$ as in Eq.~(\ref{eq:Gamma_RI_fermion}); this is recovered by combining the exact map (\ref{eq:exact_affine_recurrence}) with the short-time expansion of $b(\tau)$ (cf.\ Appendix~B). Equations~(\ref{eq:exact_affine_recurrence})-(\ref{eq:solution_affine}) reduce the RI problem to the computation of the single propagator element $U_{00}(\tau)$ (and, if needed, $b(\tau)$) from the microscopic one-particle Hamiltonian $M$. 

\begin{figure*}[t]
\centering
\subfigure[]{\includegraphics[width=0.48\textwidth]{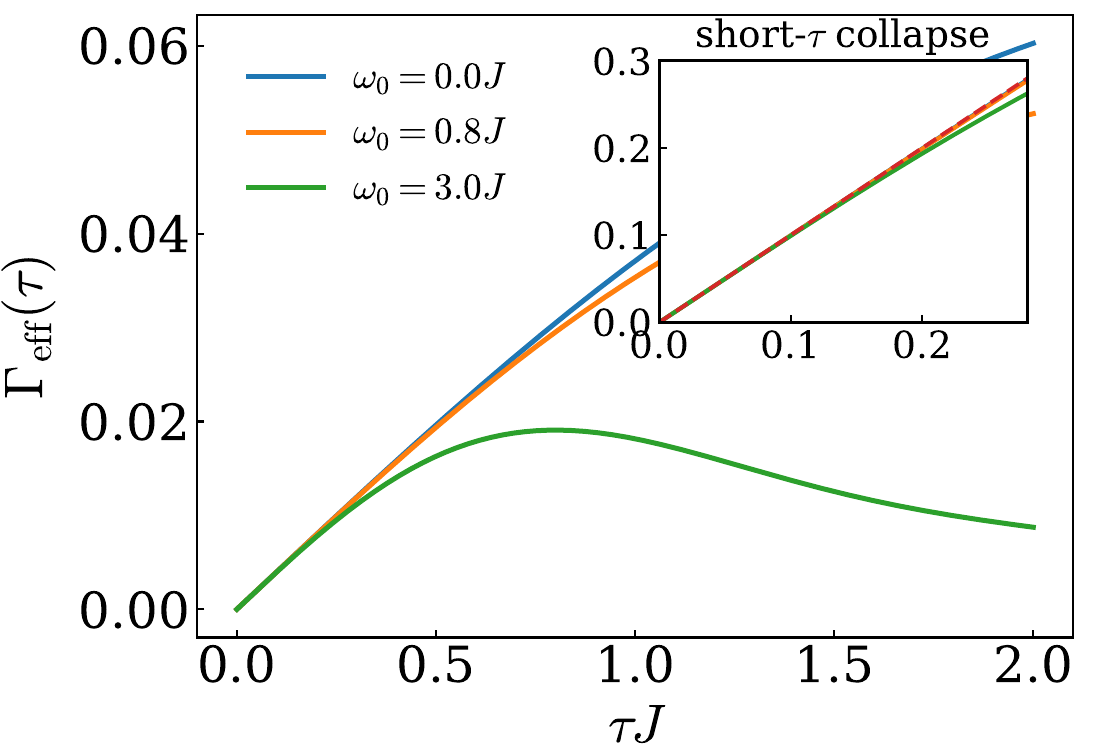}} 
\subfigure[]{\includegraphics[width=0.48\textwidth]{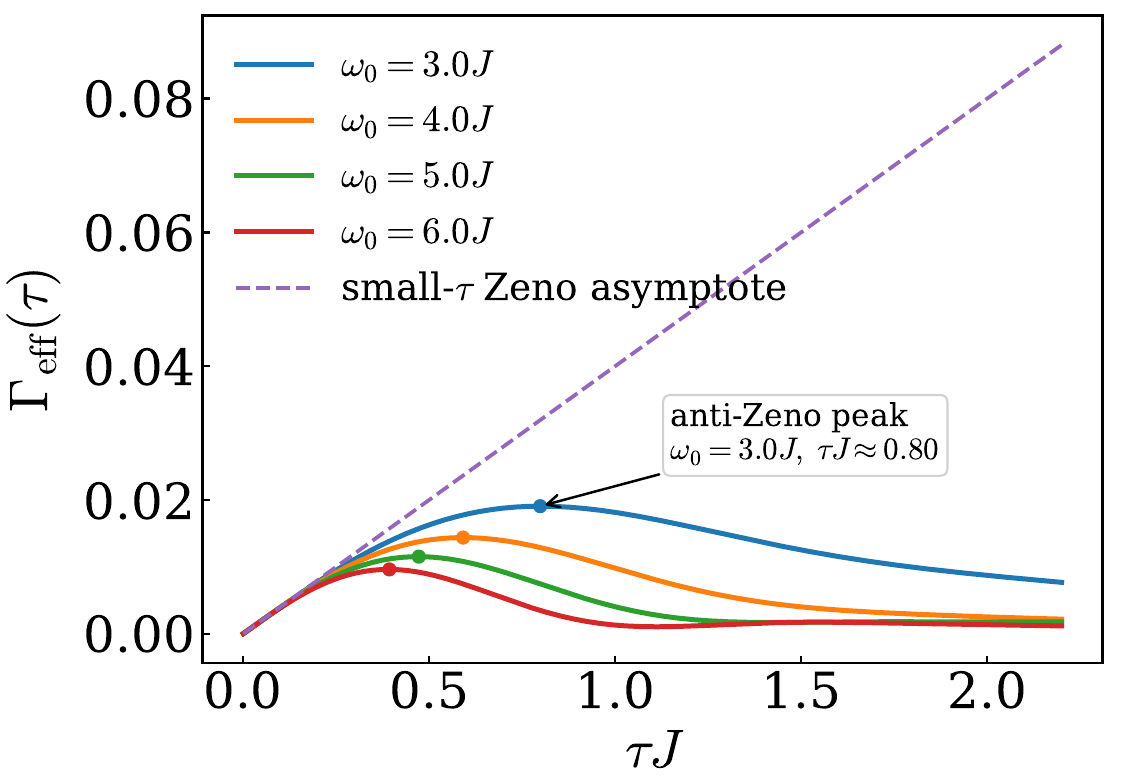}}
\caption{(a) Exact repeated-interaction (RI) decay rate for a single fermionic level coupled to a semi-infinite tight-binding chain, illustrating strict Zeno freezing. The plotted quantity is the exact map-extracted rate \(\Gamma_{\mathrm{eff}}(\tau)=-\tau^{-1}\ln |U_{00}(\tau)|^2,\) shown as a function of the reset interval $\tau J$ for three level positions: $\omega_0=0$, $\omega_0=0.8J$, and $\omega_0=3J$. In all cases the rate vanishes as $\tau\to0$, demonstrating strict Zeno suppression in the RI protocol. The inset shows the universal short-time collapse predicted by Sec.~III: after normalization by $A_T J$, with $A_T=\sum_{x\in E}|M_{0x}|^2=t_{\rm c}^2$ for the present site-basis tight-binding model, the curves collapse onto the linear law $\Gamma_{\mathrm{eff}}(\tau)/(A_TJ)=\tau J$. Numerical parameters: $J=1$, $t_{\rm c}=0.2$, truncated bath size $N_{\rm b}=400$. (b) Exact finite-$\tau$ anti-Zeno behavior. The exact map-extracted decay rate \(\Gamma_{\mathrm{eff}}(\tau)\) is shown for outside-band level positions $\omega_0=3J$, $4J$, $5J$, and $6J$, i.e. for levels lying above the tight-binding bath band $[-2J,2J]$. The dashed line is the universal small-$\tau$ Zeno asymptote, \(\Gamma_{\mathrm{eff}}(\tau)\simeq t_{\rm c}^2 \tau,\) valid in the strict frequent-reset regime. All curves follow this Zeno law as $\tau\to0$, but at finite reset interval they develop nonmonotonic structure. The marked point identifies the clearest anti-Zeno peak.}
\label{fig:zeno_aze}
\end{figure*}

\subsection{Anti-Zeno regime in the RI protocol}
\label{sec:RI_antiZeno_discussion}

In the strict frequent-reset limit, the dynamics is universally Zeno-like: the reset erases all system--environment coherences after every step, so the system is forced back to the same short-time quadratic regime at the beginning of each cycle. The possibility of an anti-Zeno effect, therefore, does \emph{not} arise inside the universal short-time quadratic regime itself. Instead, it appears only at finite reset interval $\tau$, once the evolution over a single cycle is no longer controlled solely by the leading $\tau^2$ expansion. In the exact-map formulation of Sec.~III, the survival probability obeys the Eq. (\ref{eq:exact_affine_recurrence}) with exact effective decay rate given by Eq. (\ref{eq:gamma_decay_map}). Thus the finite-$\tau$ structure of the RI dynamics is entirely encoded in the exact one-cycle survival
amplitude
\begin{equation}
    U_{00}(\tau)=\langle 0|e^{-iM\tau}|0\rangle.
    \label{eq:one-cycle-survival}
\end{equation}
While the limit $\tau\to0$ is universally controlled by the quadratic expansion of $U_{00}(\tau)$,
the behavior at larger $\tau$ depends on the full spectrum and eigenvector structure of the microscopic
single-particle Hamiltonian $M$.

This makes the physical origin of the anti-Zeno effect in RI particularly transparent.
In the exact-map language, anti-Zeno behavior is not a consequence of the short-time expansion,
but of the nontrivial finite-$\tau$ interference structure of $U_{00}(\tau)$.
Writing
\begin{equation}
    U_{00}(\tau)=\sum_{\mu} |\langle \mu|0\rangle|^2 e^{-i\varepsilon_\mu \tau},
\end{equation}
where $M|\mu\rangle=\varepsilon_\mu|\mu\rangle$, we see that the one-cycle survival amplitude is a
coherent sum of phase factors weighted by the overlap of the initial system level with the exact
single-particle eigenmodes.
For very small $\tau$, all phases remain close to unity and the universal Zeno law is recovered.
At finite $\tau$, however, these phases can interfere destructively, suppressing $|U_{00}(\tau)|$
more strongly than in the strict-Zeno regime and thereby increasing (\ref{eq:gamma_decay_map})

This is the microscopic origin of the RI anti-Zeno window. In this setting it is useful to distinguish two notions. First, there is the \emph{strict Zeno limit}, defined by the asymptotic behavior as $\tau\to0$.
For the RI protocol this limit is always Zeno, because $\Gamma_{\mathrm{RI}}(\tau)\to0$ linearly. Second, there is the \emph{finite-$\tau$ anti-Zeno regime}, defined operationally as an interval of reset times for which the exact map-extracted decay rate increases as one moves away from the strict-Zeno regime. Equivalently, one may say that an anti-Zeno window is present if there exists a finite interval of $\tau$ over which $\Gamma_{\mathrm{eff}}(\tau)$ is enhanced relative to its small-$\tau$ Zeno trend.

For the structured tight-binding environment considered later, this distinction is essential. The semi-infinite chain has a finite bandwidth and a nontrivial local density of states \cite{harrison1989electronic,grosso2014solid,colombo2005tight,zou2026review,pettifor1989tight}, so the exact survival amplitude $U_{00}(\tau)$ acquires oscillatory structure that depends sensitively on the location of the system level relative to the band. When the system level lies well inside the band, the exact RI rate may remain mostly monotone, showing a smooth crossover from strict Zeno suppression at small $\tau$ to ordinary finite-rate decay. By contrast, near the band edge or outside the band, the survival amplitude develops pronounced finite-$\tau$ oscillations, and the exact decay rate can become nonmonotonic as a function of the reset interval. This nonmonotonic regime is precisely the anti-Zeno regime of the RI protocol.

It is important to stress that this does not contradict the strict Zeno law. There is no anti-Zeno effect in the asymptotic $\tau\to0$ region itself. Rather, the RI protocol supports anti-Zeno behavior only \emph{after} the system leaves the universal quadratic regime and begins to probe the detailed spectral and interference structure of the microscopic Hamiltonian over a single reset interval. The same reset rule therefore generates both phenomena:
strict freezing in the ideal frequent-reset limit, and anti-Zeno enhancement at finite $\tau$ once the exact one-cycle propagator deviates sufficiently from its short-time expansion.

\subsubsection{Single level coupled to a semi-infinite tight-binding chain: finite-$\tau$ anti-Zeno window}
\label{sec:tb_antizeno}

To exhibit a genuine anti-Zeno regime one must go beyond the universal short-time expansion. The results of the past subsections imply that in the repeated-interaction (RI) protocol the strict frequent-reset limit is always Zeno-like: the per-cycle change in the survival probability is of order $\tau^2$, so that the effective decay rate scales linearly with $\tau$,
\begin{equation}
    \Gamma_{\mathrm{RI}}(\tau)=A_T \tau + O(\tau^2),
\end{equation}
and therefore vanishes as $\tau\to0$.
Thus the short-time theory by itself can only demonstrate strict Zeno freezing, not a finite-$\tau$ anti-Zeno enhancement.

A concrete microscopic realization in which the finite-$\tau$ structure can be computed exactly is provided by a single fermionic level of energy $\omega_0$ coupled to a semi-infinite tight-binding chain acting as a structured bath. The one-particle Hamiltonian is \cite{hoekstra1988tight,sedlmayr2013,pettifor1989tight,economou2006greens}
\begin{align}
    M=
\omega_0 |0\rangle\langle 0|
&-J\sum_{j=1}^{\infty}\Bigl(|j\rangle\langle j+1|+|j+1\rangle\langle j|\Bigr) \nonumber\\
&+t_{\rm c}\Bigl(|0\rangle\langle 1|+|1\rangle\langle 0|\Bigr),
\end{align}
where $|0\rangle$ denotes the system level and $|j\rangle$, $j\ge1$, label the bath sites.
For numerical evaluation we truncate the chain to a large number of sites $N_{\rm b}$ and compute the exact one-cycle propagator \(U(\tau)=e^{-iM\tau}\).

For an initially empty fermionic bath, the system occupation obeys the exact affine map
\begin{equation}
    P_{n+1}=a(\tau)P_n;
\qquad
a(\tau)=|U_{00}(\tau)|^2,
\end{equation}
so that the exact finite-$\tau$ decay rate is given by Eq. (\ref{eq:gamma_decay_map}). This expression is fully nonperturbative in both the coupling and the reset interval and is therefore a filter-free characterization of the RI dynamics.

For very small $\tau$ we recover the universal Zeno law,
\begin{equation}
    \Gamma_{\mathrm{eff}}(\tau)\simeq \tau\sum_{x\in E}|M_{0x}|^2.
\end{equation}
In the present site-basis tight-binding model only the first bath site couples directly to the system, so
\begin{equation}
    \sum_{x\in E}|M_{0x}|^2=t_{\rm c}^2,
\end{equation}
and the small-$\tau$ asymptote is simply
\begin{equation}
    \Gamma_{\mathrm{eff}}(\tau)\simeq t_{\rm c}^2\,\tau.
\end{equation}

The anti-Zeno effect appears only once the reset interval leaves this universal short-time regime.
At finite $\tau$, the exact survival amplitude given by Eq. (\ref{eq:one-cycle-survival}) contains the full interference structure generated by the bath spectrum. For levels near or outside the tight-binding band, this produces a nonmonotonic $\Gamma_{\mathrm{eff}}(\tau)$ with a finite-$\tau$ maximum. This finite-$\tau$ enhancement defines the anti-Zeno window of the RI protocol. It does not contradict the strict Zeno law, because the latter applies only asymptotically as $\tau\to0$, whereas the anti-Zeno regime is an intrinsically finite-$\tau$ interference phenomenon.

Figure~\ref{fig:zeno_aze}(a) shows the exact RI rate for the tight-binding example. All curves vanish as $\tau\to0$, directly demonstrating strict Zeno freezing. The inset confirms the universal short-time law of Sec.~III: after normalization by $A_TJ$, the curves collapse onto the linear relation $\Gamma_{\mathrm{eff}}(\tau)/(A_TJ)=\tau J$. Figure~\ref{fig:zeno_aze}(b) shows that the RI protocol can nevertheless exhibit a finite-$\tau$ anti-Zeno regime. For outside-band level positions, the exact map-extracted rate is nonmonotonic: after initially following the universal Zeno asymptote, it reaches a finite-$\tau$ maximum before decreasing again. This finite-$\tau$ maximum is the anti-Zeno window.
\begin{figure}[t]
\centering
\includegraphics[width=0.48\textwidth]{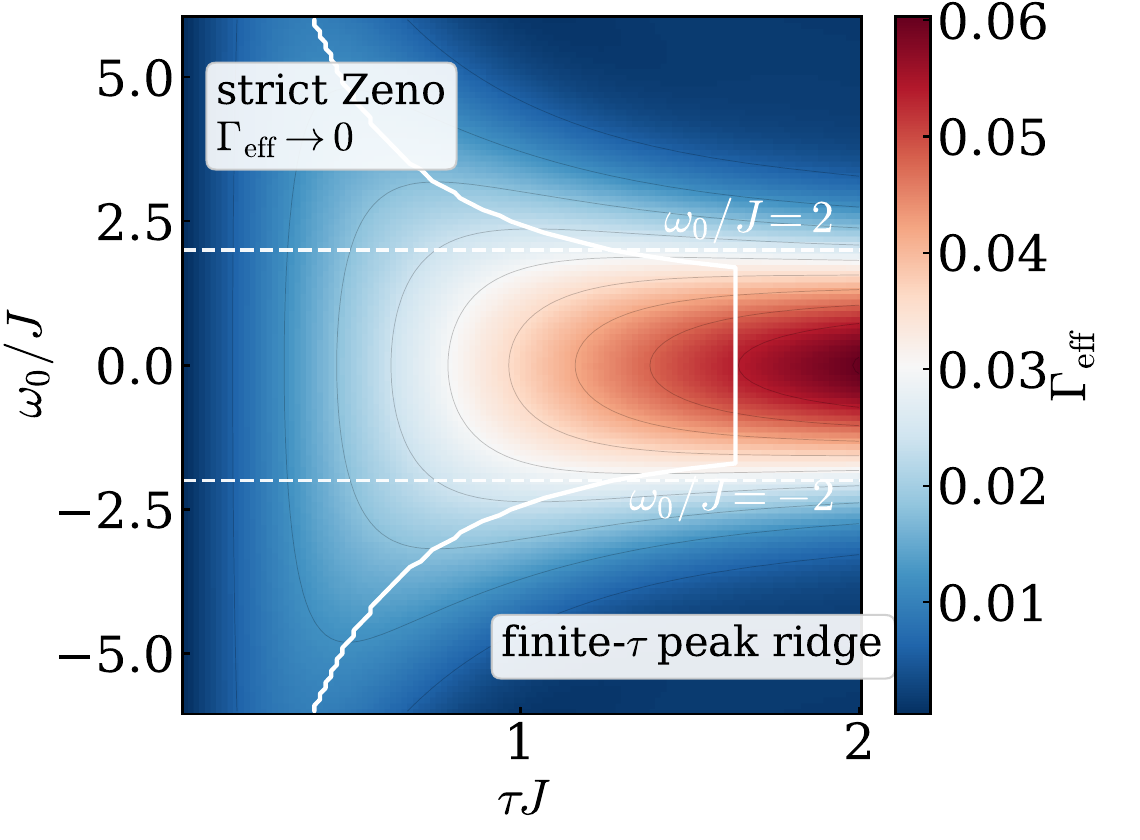} 
\caption{\textit{Dynamical design map for the repeated-interaction (RI) protocol.} The plotted quantity is the exact map-extracted decay rate \(\Gamma_{\mathrm{eff}}(\tau,\omega_0) = -\frac{1}{\tau}\ln |U_{00}(\tau)|^2,\) shown over the parameter plane \((\tau J,\omega_0/J)\). The dashed horizontal lines mark the tight-binding bath band edges, \(\omega_0/J=\pm2\). The dark region near \(\tau\to0\) corresponds to strict Zeno suppression, since \(\Gamma_{\mathrm{eff}}(\tau)\to0\) in the frequent-reset limit. The white curve is a numerical ridge of finite-\(\tau\) local maxima of \(\Gamma_{\mathrm{eff}}\), indicating where anti-Zeno enhancement is strongest. Thus the same exact RI map displays both the strict Zeno regime and the finite-\(\tau\) anti-Zeno sector within a single parameter-space representation. Numerical parameters: \(J=1\), \(t_{\rm c}=0.2\), \(\tau J\in[0.01,2.0]\), \(\omega_0/J\in[-6,6]\). The semi-infinite bath is represented numerically by a large truncated open chain.}
\label{fig:money}
\end{figure}

Figure~\ref{fig:money} summarizes the repeated-interaction protocol as a two-parameter design map in the plane \((\tau J,\omega_0/J)\). Figure~\ref{fig:money} shows the exact dynamical landscape \(\Gamma_{\mathrm{eff}}(\tau,\omega_0) = -\frac{1}{\tau}\ln|U_{00}(\tau)|^2,\) obtained directly from the exact one-cycle propagator of the single-level plus tight-binding-chain model. In this representation, the strict frequent-reset regime appears as the dark region near \(\tau\to0\), where the exact RI decay rate vanishes and the dynamics is Zeno-suppressed. The dashed horizontal lines at \(\omega_0/J=\pm2\) mark the bath band edges. Inside this strip the system level lies in the bath continuum, whereas outside it the level is spectrally detached from the band. The white curve is not a phase boundary in the thermodynamic sense, but a numerical ridge of finite-\(\tau\) local maxima of \(\Gamma_{\mathrm{eff}}\). It therefore identifies the locus of strongest anti-Zeno enhancement. In this way the figure provides a regime map of the RI dynamics: strict Zeno freezing as \(\tau\to0\), and finite-\(\tau\) anti-Zeno structure emerging once the exact one-cycle propagator probes the detailed spectral and interference properties of the structured bath.

\section{Evolving-correlation protocol and absence of strict Zeno freezing}
\label{sec:EC}

We now turn to the EC protocol, in which the environment block $\rho_{EE}$ is reset to a fixed reference state after each unitary step, but the system-environment coherences are left intact. The stroboscopic map~\eqref{eq:stroboscopic_map} applies as before, but the kept vector $V$ now contains $\rho_{SS}$ and $\rho_{SE},\rho_{ES}$. In the limit of frequent resetting ($\tau\to 0$) with fixed total time $t$, we can approximate the map by a continuous linear evolution. Expanding $D(\tau)$ and $C(\tau)$ to first order in $\tau$ yields
\begin{align}
    D(\tau) &= I + \tau\,\mathsf{D} + O(\tau^2);\\
    C(\tau) &= \tau\,\mathsf{C} + O(\tau^2),
\end{align}
and Eq.~\eqref{eq:stroboscopic_map} becomes, to leading order,
\begin{equation}
    \frac{V[n+1]-V[n]}{\tau} = \mathsf{D}V[n] + \mathsf{C} + O(\tau).
\end{equation}
In the limit $\tau\to 0$ we thus obtain the continuous-reset equation
\begin{equation}
    \frac{dV}{dt} = \mathsf{D}\,V + \mathsf{C}.
\end{equation}
In the EC protocol the environment block of the SPDM is reset to a fixed reference matrix $\rho_{EE}^{(0)}$ at every stroboscopic step, but the system--environment coherences $\rho_{SE}$ and $\rho_{ES}$ are retained. In the continuous-reset limit discussed in Sec.~\ref{subsec:stroboscopic_map}, this yields a linear differential equation for the kept SPDM blocks. Writing
\begin{equation}
  \rho(t) =
  \begin{pmatrix}
    \rho_{SS}(t) & \rho_{SE}(t)\\
    \rho_{ES}(t) & \rho_{EE}(t)
  \end{pmatrix};
  \,\,\,
  M =
  \begin{pmatrix}
    M_{SS} & M_{SE}\\
    M_{ES} & M_{EE}
  \end{pmatrix},
\end{equation}
the generator of the continuous EC dynamics is obtained by replacing $\rho_{EE}(t)$ by the fixed reference block $\rho_{EE}^{(0)}$ in the unitary equation $\dot{\rho}=-i[M,\rho]$, while keeping $\rho_{SS}$ and $\rho_{SE},\rho_{ES}$ as dynamical variables. This gives the block equations
\begin{align}
  \dot{\rho}_{SS}
  &= -i[M_{SS},\rho_{SS}]
     -i\big(M_{SE}\rho_{ES}-\rho_{SE}M_{ES}\big);
  \label{eq:EC_block_SS_main}\\
  \dot{\rho}_{SE}
  &= -i\big(
        M_{SS}\rho_{SE}+M_{SE}\rho_{EE}^{(0)}\notag\\
        &-\rho_{SS}M_{SE}-\rho_{SE}M_{EE}
      \big),
  \label{eq:EC_block_SE_main}
\end{align}
with $\rho_{ES}=\rho_{SE}^\dagger$.

To analyze the survival probability in direct analogy with the RI case, we now specialize to a single fermionic level in the system, labelled by $0\in S$, coupled to an environment that has been diagonalized in the eigenbasis of $M_{EE}$. We denote the occupation of the system level by
\begin{equation}
  P(t) \equiv \rho_{00}(t),
\end{equation}
and write the system-environment coherences as
\begin{equation}
  s_k(t) \equiv \rho_{0k}(t);\qquad
  \rho_{k0}(t)=s_k^*(t),
\end{equation}
where $k$ labels environment eigenmodes with single-particle energies $\omega_k$ and couplings $g_k\equiv M_{0k}$. The reference environment state is taken to be diagonal in this eigenbasis with occupations $(\rho_{EE}^{(0)})_{kk}=n_k$.

Inserting these definitions into Eqs.~\eqref{eq:EC_block_SS_main}-\eqref{eq:EC_block_SE_main}, and using that
$M_{SS}=\omega_0$ is a scalar, we obtain
\begin{align}
  \dot{P}(t)
  &= -i\sum_k\big(g_k s_k^*(t) - s_k(t)g_k^*\big); \label{eq:EC_P_dot_main}\\
  \dot{s}_k(t)
  &= -i\big[\Delta_k s_k(t) + g_k\big(n_k-P(t)\big)\big],
  \label{eq:EC_sk_dot_main}
\end{align}
where $\Delta_k\equiv\omega_0-\omega_k$. Equation~\eqref{eq:EC_sk_dot_main} is a linear inhomogeneous equation for the coherences $s_k(t)$. Assuming $s_k(0)=0$ for all $k$, we solve (\ref{eq:EC_sk_dot_main}) by the integrating-factor method. Multiply both sides by $e^{+i\Delta_k t}$:
\begin{equation}
    \frac{d}{dt}\Bigl(e^{+i\Delta_k t}s_k(t)\Bigr)
=
-i g_k\,e^{+i\Delta_k t}\,[n_k-P(t)].
\end{equation}
Integrating from $0$ to $t$ gives the formal solution
\begin{equation}
s_k(t)
=
-i g_k\int_0^t dt'\,e^{-i\Delta_k(t-t')}\,[n_k-P(t')],
\end{equation}
and hence $s_k^\ast(t)=+i g_k^\ast\int_0^t dt'\,e^{+i\Delta_k(t-t')}\,[n_k-P(t')]$. Substituting $s_k$ and $s_k^*$ into Eq.~\eqref{eq:EC_P_dot_main} yields an integro-differential equation for the survival probability,
\begin{align}
  \dot{P}(t)
  = -2\sum_k |g_k|^2 \int_0^t dt'\,
      &\cos\big[\Delta_k(t-t')\big]\notag\\ &\times \big(P(t')-n_k\big).
  \label{eq:EC_P_memory_main}
\end{align}

Introducing the spectral density $J(\omega)=\sum_k |g_k|^2\delta(\omega-\omega_k)$ and the occupation function $n(\omega)$ defined by $n(\omega_k)=n_k$, Eq.~\eqref{eq:EC_P_memory_main} can be written in continuum form as
\begin{align}
  \dot{P}(t)
  = -2\int d\omega\,J(\omega)\int_0^t dt'\,
      &\cos\big[(\omega_0-\omega)(t-t')\big]\notag\\
      &\times\big(P(t')-n(\omega)\big).
  \label{eq:EC_P_memory_cont_main}
\end{align}
Under standard weak-coupling and smooth-spectrum assumptions (Markov approximation), one may approximate $P(t')\simeq P(t)$ inside the integral and extend the upper limit of $\int_0^{t}d(t-t')$ to infinity, obtaining
\begin{equation}
  \int_0^\infty d\tau\,\cos\big[(\omega_0-\omega)\tau\big]
  = \pi\delta(\omega_0-\omega),
\end{equation}
where principal-value terms generate only a Lamb shift that we neglect here. Equation~\eqref{eq:EC_P_memory_cont_main} then reduces to the Markovian rate equation
\begin{equation}
  \dot{P}(t)
  = -\Gamma_{\mathrm{EC}}\big(P(t)-n(\omega_0)\big);
  \qquad
  \Gamma_{\mathrm{EC}} = 2\pi J(\omega_0).
  \label{eq:EC_rate_equation_main}
\end{equation}
Thus the survival probability relaxes exponentially to the bath occupation $n(\omega_0)$ with a decay rate
$\Gamma_{\mathrm{EC}}$ that depends only on the local spectral density $J(\omega_0)$, and \emph{not} on the reset interval $\tau$ in the continuous-reset limit.

Comparing Eq.~\eqref{eq:EC_rate_equation_main} with the RI result $\Gamma_{\mathrm{RI}}(\tau)\propto\tau$ for $\tau\to 0$, we see that the EC protocol does \emph{not} exhibit strict Zeno freezing: even as the stroboscopic reset is made arbitrarily frequent, the effective decay rate remains finite and equal to the usual Fermi golden-rule value $2\pi J(\omega_0)$. Physically, this difference arises because the EC protocol continually re-injects system-environment coherences via the fixed environment block
$\rho_{EE}^{(0)}$, so that the short-time quadratic suppression of decay that underlies the Zeno effect is not restored by the reset. In contrast, the RI protocol erases $\rho_{SE}$ and $\rho_{ES}$ at every step, forcing the dynamics back into the universal quadratic short-time regime and producing the linear $\Gamma_{\mathrm{RI}}(\tau)\propto\tau$ behavior.

In other words, unlike the RI protocol where system-environment coherences are erased at each step and the per-cycle change in the system block is of order $\tau^2$, the EC protocol retains these coherences and yields an $O(\tau)$ change per cycle in the system block, which translates into an $O(1)$ drift in the continuous limit. Therefore, strict Zeno freezing (vanishing decay rate as $\tau\to 0$) does \emph{not} occur in EC.

\section{Discussion and conclusions}

We have analyzed quantum Zeno and anti-Zeno dynamics in a broad class of quadratic open quantum systems subject to stroboscopic resetting of the environment degrees of freedom. Working directly at the level of the single-particle density matrix, we derived the exact stroboscopic affine map for the kept correlators and identified two natural reset protocols, RI and EC, distinguished by whether system--environment coherences are erased or preserved after each reset step.

For the repeated-interaction protocol, we derived an explicit short-time form of the Zeno law for the survival probability of a single-particle excitation, valid for a single system level quadratically coupled to an arbitrary Gaussian environment. We showed that the per-cycle change in the relevant population is of order $\tau^2$, which implies an effective decay rate of the form $\Gamma_{\rm RI}(\tau)\propto \tau$ and therefore strict Zeno freezing in the limit of infinitely frequent resetting. Beyond this universal short-time regime, we formulated the RI dynamics directly in terms of the exact one-cycle propagator. In particular, for the single-level problem the effective decay rate is determined by the exact survival amplitude $U_{00}(\tau)$, so that the full finite-$\tau$ behavior follows from the exact map itself rather than from an additional approximation scheme. This exact-map formulation makes transparent how structured environments generate nonmonotonic finite-$\tau$ behavior and, consequently, anti-Zeno windows. The tight-binding example illustrates this explicitly: while all curves exhibit strict Zeno suppression as $\tau\to0$, levels near or outside the bath band develop finite-$\tau$ maxima in the exact decay rate.

For the evolving-correlation protocol, we obtained the continuous-reset generator for the coupled dynamics of the system block and the retained system-environment coherences. In contrast to RI, the reset does not erase these coherences after each step. As a result, the inhomogeneous terms generated by the environment reset drive the coherences to finite values on the system time scale, and these feed back into the system block already in the $\tau\to0$ limit. Consequently, strict Zeno freezing does not occur in EC. The EC protocol can still display Zeno-like slowing down and anti-Zeno enhancement at finite $\tau$, but its asymptotic frequent-reset limit is fundamentally different from that of RI. This sharp contrast shows that the presence or absence of retained system-environment coherences is a central organizing principle for Zeno physics in reset-based quadratic open-system dynamics. The framework connects naturally to the literature on quantum Zeno dynamics~\cite{MisraSudarshan1977,FacchiNakazatoPascazio2001,Becker2021}, anti-Zeno phenomena~\cite{KofmanKurizki2000Nature,Ruseckas2004}, and quantum collision models~\cite{Ciccarello2022PhysRep,Cusumano2022Entropy}. Possible extensions include the incorporation of explicitly non-Markovian reset rules, the study of interacting systems perturbatively beyond the quadratic limit, and applications to engineered reservoir design and quantum thermodynamic cycles in which reset protocols are used to control both dynamical slowdown or enhancement and their associated thermodynamic cost.

\section*{Data availability}

All data shown in the figures are available from the authors upon reasonable request.
\medskip

\section{Acknowledgements}

J.K. and T.A-N. have been supported by the Academy of Finland through its QTF Center of Excellence grant no. 312298 and by the European Union and the European Innovation Council through the Horizon Europe project QRC-4-ESP (grant No. 101129663), and EU Horizon Europe Quest project (No. 10116088). A.L. acknowledges support from the Leverhulme Trust Research Project Grant RPG-2025-063. 

\appendix

\section{Explicit evaluation of the double commutator}
\label{app:double_commutator}

Here we derive Eq.~\eqref{eq:double_comm_identity}, which we reproduce here:
\begin{equation}
    [M,[M,\rho]]_{00} = 2\sum_{x\in E}|M_{0x}|^2\big(\rho_{00}-n_x\big).
\end{equation}
We start from the general definition
\begin{equation}
    [M,\rho]_{ab} = \sum_c \big(M_{ac}\rho_{cb} - \rho_{ac}M_{cb}\big),
    \label{eq:app_comm_def}
\end{equation}
and
\begin{equation}
    [M,[M,\rho]]_{00}
    = \sum_\gamma \Big(M_{0\gamma}[M,\rho]_{\gamma 0} - [M,\rho]_{0\gamma}M_{\gamma 0}\Big).
    \label{eq:app_double_comm_start}
\end{equation}

At the beginning of each unitary step in the RI protocol we assume:
\begin{enumerate}
    \item The environment block $\rho_{EE}$ is diagonal in the eigenbasis of $M_{EE}$:
    $\rho_{ee'} = n_e\,\delta_{ee'}$, $e,e'\in E$.
    \item The system-environment coherences vanish: $\rho_{Se}=\rho_{eS}=0$.
    \item The system block $\rho_{SS}$ is diagonal, with system level labelled by $0$:
    $\rho_{00}$ is the system occupation we are interested in.
\end{enumerate}
Thus, in the combined index set $S\cup E$ we can write
\begin{equation}
    \rho_{ab} = \rho_a\,\delta_{ab}, \qquad
    \rho_a =
    \begin{cases}
        \rho_{00}, & a = 0;\\[2pt]
        n_a, & a\in E.
    \end{cases}
    \label{eq:app_rho_diag}
\end{equation}

Using Eq.~\eqref{eq:app_comm_def} and the diagonal form~\eqref{eq:app_rho_diag}, we obtain a very simple expression for the commutator:
\begin{align}
    [M,\rho]_{ab}
    &= \sum_c \big(M_{ac}\rho_{cb} - \rho_{ac}M_{cb}\big)\notag\\
    &= \sum_c \big(M_{ac}\rho_c\delta_{cb} - \rho_a\delta_{ac}M_{cb}\big)\notag\\
    &= M_{ab}\rho_b - \rho_a M_{ab}\notag\\
    &= \big(\rho_b - \rho_a\big)M_{ab}.
    \label{eq:app_comm_diag}
\end{align}

We now evaluate the two matrix elements appearing in
Eq.~\eqref{eq:app_double_comm_start}. For $\gamma$ arbitrary,
\begin{align}
    [M,\rho]_{\gamma 0}
    &= \big(\rho_0 - \rho_\gamma\big)M_{\gamma 0};\label{eq:app_comm_gamma0}\\
    [M,\rho]_{0\gamma}
    &= \big(\rho_\gamma - \rho_0\big)M_{0\gamma}.
    \label{eq:app_comm_0gamma}
\end{align}
Substituting Eqs.~\eqref{eq:app_comm_gamma0} and
\eqref{eq:app_comm_0gamma} into Eq.~\eqref{eq:app_double_comm_start}
gives
\begin{align}
    [M,[M,\rho]]_{00}
    &= \sum_\gamma \Big(
        M_{0\gamma}\big(\rho_0 - \rho_\gamma\big)M_{\gamma 0}\notag\\
        &- \big(\rho_\gamma - \rho_0\big)M_{0\gamma}M_{\gamma 0}
      \Big)\notag\\
    &= \sum_\gamma \Big(
        (\rho_0 - \rho_\gamma)M_{0\gamma}M_{\gamma 0}\notag\\
        &+ (\rho_0 - \rho_\gamma)M_{0\gamma}M_{\gamma 0}
      \Big)\notag\\
    &= 2\sum_\gamma (\rho_0 - \rho_\gamma)\,|M_{0\gamma}|^2.
    \label{eq:app_double_comm_general}
\end{align}
The term with $\gamma=0$ vanishes identically because
$\rho_0 - \rho_0 = 0$, so only environment indices contribute.
Writing $\rho_0 = \rho_{00}$ and $\rho_x = n_x$ for $x\in E$, we obtain
\begin{equation}
    [M,[M,\rho]]_{00}
    = 2 \sum_{x\in E} |M_{0x}|^2\big(\rho_{00} - n_x\big),
\end{equation}
which is precisely Eq.~\eqref{eq:double_comm_identity}.

\section{Details of the short-time expansion and decay rates}
\label{app:short_time_details}

We give a step–by–step derivation of the
short-time expansion used in Sec.~III and of the Zeno decay
rates quoted in Eqs.~(\ref{eq:Gamma_RI_small_tau})–(\ref{eq:Gamma_RI_fermion}).  We work entirely at the
single-particle density-matrix (SPDM) level.

\subsection{Check of Eq.~(\ref{eq:rho_prime_commutator}): BCH expansion}
\label{app:BCH_check}

The single-particle Hamiltonian is the Hermitian matrix $M$,
and the unitary propagator for a short time $\tau$ is
\begin{equation}
  U(\tau) = e^{-iM\tau},\qquad U^\dagger(\tau)=e^{+iM\tau}.
\end{equation}
Expanding both to second order in $\tau$ gives
\begin{align}
  U(\tau)
  &= I - iM\tau - \frac{1}{2}M^2\tau^2 + O(\tau^3); \label{eq:U_exp_app}\\
  U^\dagger(\tau)
  &= I + iM\tau - \frac{1}{2}M^2\tau^2 + O(\tau^3), \label{eq:Ud_exp_app}
\end{align}
where $I$ is the identity matrix.

The SPDM after one unitary step is
\begin{equation}
  \rho' = U^\dagger(\tau)\,\rho\,U(\tau).
\end{equation}
We now expand this product explicitly up to $O(\tau^2)$.
First multiply on the right:
\begin{align}
  \rho\,U(\tau)
  &= \rho\Big(I - iM\tau - \tfrac{1}{2}M^2\tau^2\Big) + O(\tau^3)\notag\\
  &= \rho - i\rho M\tau - \frac{1}{2}\rho M^2\tau^2 + O(\tau^3).
  \label{eq:rhoU_app}
\end{align}
Next multiply on the left by $U^\dagger(\tau)$:
\begin{align}
  \rho'
  &= \Big(I + iM\tau - \tfrac{1}{2}M^2\tau^2\Big)
     \Big(\rho - i\rho M\tau - \tfrac{1}{2}\rho M^2\tau^2\Big)\notag\\
     &+ O(\tau^3). \label{eq:Ud_rhoU_start}
\end{align}
We now expand term by term.

\emph{(i) Terms proportional to $I$:}
\begin{equation}
  I \cdot (\rho - i\rho M\tau - \tfrac{1}{2}\rho M^2\tau^2)
  = \rho - i\rho M\tau - \tfrac{1}{2}\rho M^2\tau^2.
  \label{eq:I_block}
\end{equation}

\emph{(ii) Terms proportional to $iM\tau$:}
\begin{align}
  iM\tau\cdot(\rho - i\rho M\tau)
  &= iM\rho\tau + (iM\tau)(-i\rho M\tau) + O(\tau^3)\notag\\
  &= iM\rho\tau + (-i^2)M\rho M\tau^2 + O(\tau^3)\notag\\
  &= iM\rho\tau + M\rho M\tau^2 + O(\tau^3).
  \label{eq:iM_block}
\end{align}
(The product with $-\frac{1}{2}\rho M^2\tau^2$ is of order $\tau^3$
and is thus discarded.)

\emph{(iii) Terms proportional to $-\frac{1}{2}M^2\tau^2$:}
\begin{align}
  -\tfrac{1}{2}M^2\tau^2\cdot\rho
  &= -\tfrac{1}{2}M^2\rho\tau^2,
  \label{eq:M2_block}
\end{align}
while the products with $-i\rho M\tau$ and $-\tfrac{1}{2}\rho M^2\tau^2$
would give $\tau^3$ and $\tau^4$ contributions and are therefore neglected.

Collecting all contributions
\eqref{eq:I_block}–\eqref{eq:M2_block}, we obtain
\begin{align}
  \rho'
  &= \rho
   - i\rho M\tau
   - \tfrac{1}{2}\rho M^2\tau^2
   + iM\rho\tau \notag\\
   &+ M\rho M\tau^2
   - \tfrac{1}{2}M^2\rho\tau^2
   + O(\tau^3)\notag\\[2pt]
  &= \rho
   + i(M\rho - \rho M)\tau \notag\\
   &+ \Big(M\rho M - \tfrac{1}{2}M^2\rho - \tfrac{1}{2}\rho M^2\Big)\tau^2
   + O(\tau^3).
  \label{eq:rho_prime_pre_comm}
\end{align}

We now relate this to commutators.  By definition,
\begin{equation}
  [M,\rho] \equiv M\rho - \rho M.
\end{equation}
The double commutator is
\begin{align}
  [M,[M,\rho]]
  &= M(M\rho - \rho M) - (M\rho - \rho M)M\notag\\
  &= M^2\rho - M\rho M - M\rho M + \rho M^2\notag\\
  &= M^2\rho + \rho M^2 - 2M\rho M.
  \label{eq:double_comm_app}
\end{align}
Solving Eq.~\eqref{eq:double_comm_app} for $M\rho M - \tfrac{1}{2}M^2\rho
- \tfrac{1}{2}\rho M^2$ gives
\begin{equation}
  M\rho M - \tfrac{1}{2}M^2\rho - \tfrac{1}{2}\rho M^2
  = -\frac{1}{2}[M,[M,\rho]].
\end{equation}
Inserting this into Eq.~\eqref{eq:rho_prime_pre_comm}, we obtain
\begin{equation}
  \rho'
  = \rho
    + i[M,\rho]\tau
    - \frac{1}{2}[M,[M,\rho]]\tau^2
    + O(\tau^3),
  \label{eq:rho_BCH_appendix}
\end{equation}
which is precisely Eq.~(\ref{eq:rho_prime_commutator}) of the main text.  In particular, the signs of both the linear and quadratic terms are fixed and unambiguous.

\subsection{From Eq.~(\ref{eq:doble_commutator1}) to Eq.~(\ref{eq:Gamma_RI_small_tau}): empty fermionic bath}
\label{app:empty_bath}

We now derive the short-time Zeno law for the survival probability of a single-particle excitation in the RI protocol, making explicit the steps from Eq.~(\ref{eq:doble_commutator1}) to Eq.~(\ref{eq:Gamma_RI_small_tau}). We focus on one special system level, labelled $0\in S$, and denote its occupation at time $t_n$ by
\begin{equation}
  P_n \equiv \rho_{00}(t_n).
\end{equation}
At the beginning of the $n$th cycle (immediately after the reset) we assume:
\begin{itemize}
  \item[(i)] The SPDM is diagonal in some single-particle basis:
  \begin{equation}
    \rho_{\alpha\beta}(t_n) = \rho_\alpha(t_n)\,\delta_{\alpha\beta}.
  \end{equation}
  \item[(ii)] System–environment coherences vanish:
  \begin{equation}
    \rho_{\alpha x}(t_n) = \rho_{x\alpha}(t_n) = 0,
    \qquad \alpha\in S,\; x\in E.
  \end{equation}
  \item[(iii)] The environment block coincides with the reset state,
  \begin{equation}
    \rho_{xx}(t_n) = n_x,
    \qquad x\in E,
  \end{equation}
  with $n_x$ the occupation of mode $x$ in the reference environment.
\end{itemize}

Using the short-time expansion~\eqref{eq:rho_BCH_appendix}, the matrix element $(0,0)$ of the SPDM just before the next reset is
\begin{align}
  \rho_{00}(t_{n+1}^-)
  &= \rho_{00}(t_n)
    + i[M,\rho(t_n)]_{00}\tau \notag\\
    &- \frac{1}{2}[M,[M,\rho(t_n)]]_{00}\tau^2
    + O(\tau^3).
  \label{eq:P_before_reset}
\end{align}
For a diagonal SPDM one finds
\begin{equation}
  [M,\rho]_{\alpha\beta}
  = (\rho_\beta - \rho_\alpha)M_{\alpha\beta},
\end{equation}
so in particular
\begin{equation}
  [M,\rho]_{00}
  = (\rho_0 - \rho_0)M_{00} = 0.
\end{equation}
Hence the linear term in Eq.~\eqref{eq:P_before_reset} vanishes, and the leading short-time change of $P_n$ is controlled by the double commutator.

In Appendix~\ref{app:double_commutator} we show that for a diagonal SPDM one has
\begin{equation}
  [M,[M,\rho]]_{00}
  = 2\sum_{\gamma}|M_{0\gamma}|^2\big(\rho_{00}-\rho_{\gamma\gamma}\big),
  \label{eq:double_comm_diag_00}
\end{equation}
where the sum runs over all single-particle indices $\gamma$. In the RI setting we are interested in the exchange of population
between the distinguished system level $0$ and the environment, so we restrict the sum to environment indices $x\in E$:
\begin{equation}
  [M,[M,\rho]]_{00}
  = 2\sum_{x\in E}|M_{0x}|^2\big(\rho_{00}-n_x\big).
  \label{eq:double_comm_env_only}
\end{equation}
Substituting Eq.~\eqref{eq:double_comm_env_only} into
Eq.~\eqref{eq:P_before_reset}, we obtain
\begin{align}
  \rho_{00}(t_{n+1}^-)
  &= P_n - \frac{\tau^2}{2}\,2\sum_{x\in E}|M_{0x}|^2\big(P_n - n_x\big)
     + O(\tau^3)\notag\\
  &= P_n - \tau^2\sum_{x\in E}|M_{0x}|^2\big(P_n - n_x\big)
     + O(\tau^3).
  \label{eq:P_before_reset_expanded}
\end{align}

In the RI protocol the reset step does \emph{not} affect the system block, so
\begin{equation}
  \rho_{00}(t_{n+1}) = \rho_{00}(t_{n+1}^-) .
\end{equation}
Therefore the change in $P_n$ over one full cycle is
\begin{align}
  \Delta P
  &\equiv P_{n+1} - P_n
   = \rho_{00}(t_{n+1}) - \rho_{00}(t_n)\notag\\
  &= -\tau^2\sum_{x\in E}|M_{0x}|^2\big(P_n - n_x\big) + O(\tau^3).
  \label{eq:DeltaP_general}
\end{align}

For an empty fermionic bath we have $n_x=0$ for all $x\in E$, so
\begin{equation}
  \Delta P
  = -\tau^2\sum_{x\in E}|M_{0x}|^2P_n + O(\tau^3).
\end{equation}
Defining
\begin{equation}
  A \equiv \sum_{x\in E}|M_{0x}|^2,
  \label{eq:A_def}
\end{equation}
this becomes
\begin{equation}
  P_{n+1} = P_n - A\tau^2 P_n + O(\tau^3).
  \label{eq:Pn_update_empty}
\end{equation}

Neglecting $O(\tau^3)$ corrections, Eq.~\eqref{eq:Pn_update_empty} is a simple linear recurrence,
\begin{equation}
  P_{n+1} = (1 - A\tau^2)P_n,
\end{equation}
whose solution is
\begin{equation}
  P_n \simeq (1 - A\tau^2)^n,
  \label{eq:Pn_solution}
\end{equation}
for $n\ge 0$.

To pass to a continuous-time description we write $t=n\tau$, so that $n=t/\tau$, and rewrite Eq.~\eqref{eq:Pn_solution} as
\begin{equation}
  P(t) \simeq \big(1 - A\tau^2\big)^{t/\tau}.
\end{equation}
Using the standard logarithmic expansion $\ln(1-x) = -x + O(x^2)$ for small $x$, we have
\begin{align}
  \ln P(t)
  &= \frac{t}{\tau}\ln(1 - A\tau^2)\notag\\
  &\simeq \frac{t}{\tau}(-A\tau^2) + O(\tau^2 t)\notag\\
  &= -A\tau t + O(\tau^2 t).
\end{align}
Exponentiating, we obtain
\begin{equation}
  P(t) \simeq \exp\!\big[-A\tau t + O(\tau^2 t)\big]
           \simeq \exp\!\big[-\Gamma_{\mathrm{RI}}(\tau)t\big],
\end{equation}
where, to leading order in $\tau$,
\begin{equation}
  \Gamma_{\mathrm{RI}}(\tau) = A\tau + O(\tau^2)
  = \tau\sum_{x\in E}|M_{0x}|^2 + O(\tau^2).
  \label{eq:Gamma_RI_empty_appendix}
\end{equation}
This is Eq.~(\ref{eq:Gamma_RI_small_tau}) of the main text, and it shows explicitly that $\Gamma_{\mathrm{RI}}(\tau)\propto\tau$ as $\tau\to 0$, i.e.\ strict
Zeno freezing in the RI protocol.

\subsection{From Eq.~(\ref{eq:deltaP_RI}) to Eq.~(\ref{eq:Gamma_RI_fermion}): finite-temperature fermionic bath}
\label{app:finite_T}

We now keep the finite occupations $n_x$ of the environment modes and derive the short-time decay rate out of the initially occupied level. Starting again from Eq.~\eqref{eq:DeltaP_general}, we write
\begin{equation}
  \Delta P
  = -\tau^2\sum_{x\in E}|M_{0x}|^2\big(P_n - n_x\big) + O(\tau^3).
  \label{eq:DeltaP_finiteT_start}
\end{equation}
At $t=0$ the system level is assumed to be occupied, so $P_0\simeq 1$.  For short times (or equivalently, for a small number
of reset steps) we have $P_n\simeq 1$ up to corrections of order $\tau$, so we may write
\begin{equation}
  P_n - n_x \simeq 1 - n_x
  \qquad\text{for small $n\tau$}.
\end{equation}
Substituting this into Eq.~\eqref{eq:DeltaP_finiteT_start} gives
\begin{equation}
  \Delta P
  \simeq -\tau^2\sum_{x\in E}|M_{0x}|^2(1 - n_x).
  \label{eq:DeltaP_finiteT_approx}
\end{equation}
Defining
\begin{equation}
  A_T \equiv \sum_{x\in E}|M_{0x}|^2(1 - n_x),
  \label{eq:AT_def}
\end{equation}
we can write
\begin{equation}
  P_{n+1} - P_n \simeq -A_T\tau^2.
\end{equation}
Dividing by $\tau$ and using $t=n\tau$, we obtain to leading order
\begin{equation}
  \frac{P_{n+1} - P_n}{\tau}
  \simeq -A_T\tau,
\end{equation}
which we interpret as the short-time decay rate of the initial
occupation:
\begin{equation}
  P(t) \simeq 1 - \Gamma_{\mathrm{RI}}(\tau)t;
  \qquad
  \Gamma_{\mathrm{RI}}(\tau) = A_T\tau + O(\tau^2).
\end{equation}
Thus
\begin{equation}
  \Gamma_{\mathrm{RI}}(\tau)
  = \tau\sum_{x\in E}|M_{0x}|^2(1 - n_x) + O(\tau^2),
  \label{eq:Gamma_RI_finiteT_appendix}
\end{equation}
which is Eq.~(42) of the main text.  The factor $(1 - n_x)$ is the usual Pauli-blocking factor for decay into fermionic modes of
occupation $n_x$.  For bosonic environments one finds instead the stimulated-emission factor $(1 + n_x)$, leading to
\begin{equation}
  \Gamma_{\mathrm{RI}}^{\mathrm{(boson)}}(\tau)
  = \tau\sum_{x\in E}|M_{0x}|^2(1 + n_x) + O(\tau^2).
\end{equation}

Equations~\eqref{eq:Gamma_RI_empty_appendix} and \eqref{eq:Gamma_RI_finiteT_appendix} together provide a complete and explicit derivation of the short-time Zeno law for the RI protocol, both at zero and finite temperature.

\subsection{Short-time expansion of the affine map coefficients $a(\tau)$ and $b(\tau)$}

In Sec.~\ref{sec:model}C, the single-level RI dynamics is written as an affine map
$P_{n+1}=a(\tau)P_n+b(\tau)$ with
\begin{equation}
a(\tau)\equiv |U_{00}(\tau)|^2;
\quad
b(\tau)\equiv \sum_{e,e'\in E}U^{*}_{0e}(\tau)\,U_{0e'}(\tau)\,\rho^{(0)}_{ee'}.
\label{eqapp:ataubtau}
\end{equation}
Here $U(\tau)=e^{-iM\tau}$ is the one-particle propagator and $M$ is the Hermitian single-particle Hamiltonian matrix. We now derive the leading small-$\tau$ expansions of $a(\tau)$ and $b(\tau)$. 

\paragraph{Assumption (diagonal reset SPDM in the bath eigenbasis).}
We assume the environment reset SPDM is diagonal in some bath mode basis:
\begin{equation}
\rho^{(0)}_{ee'} = n_e\,\delta_{ee'};
\qquad e,e'\in E,
\end{equation}
with occupations $n_e$ (Fermi or Bose). Then Eq.~(B45) simplifies to
\begin{equation}
b(\tau)=\sum_{e\in E}|U_{0e}(\tau)|^2\,n_e.
\label{eqapp:btau}
\end{equation}

\paragraph{Step 1: short-time expansion of $U(\tau)$ and the matrix elements $U_{00}(\tau)$ and $U_{0e}(\tau)$.}
Using the Taylor expansion of the matrix exponential,
\begin{equation}
U(\tau)=e^{-iM\tau}
= I - iM\tau - \frac{1}{2}M^2\tau^2 + O(\tau^3),
\end{equation}
we obtain, for the diagonal element $(0,0)$,
\begin{align}
U_{00}(\tau)
&=
\bigl(I\bigr)_{00} - i(M\tau)_{00} - \frac{1}{2}(M^2\tau^2)_{00} + O(\tau^3)\nonumber\\
&=
1 - iM_{00}\tau - \frac{1}{2}(M^2)_{00}\tau^2 + O(\tau^3).
\end{align}
For an off-diagonal element $(0,e)$ with $e\in E$, we have $(I)_{0e}=0$, hence
\begin{align}
U_{0e}(\tau)
&=
0 - iM_{0e}\tau - \frac{1}{2}(M^2)_{0e}\tau^2 + O(\tau^3)\nonumber\\
&=
-iM_{0e}\tau - \frac{1}{2}(M^2)_{0e}\tau^2 + O(\tau^3).
\label{eqapp:offdiag}
\end{align}

\paragraph{Step 2: expansion of $a(\tau)=|U_{00}(\tau)|^2$.}
Define
\begin{align}
u_1&\equiv -iM_{00}; \quad u_2\equiv -\frac{1}{2}(M^2)_{00}, \nonumber\\
&\text{so that}\quad
U_{00}(\tau)=1+u_1\tau+u_2\tau^2+O(\tau^3).
\end{align}
Then
\begin{equation}
U_{00}^*(\tau)=1+u_1^*\tau+u_2^*\tau^2+O(\tau^3),
\end{equation}
and multiplying gives
\begin{align}
a(\tau)&=|U_{00}(\tau)|^2
=
(1+u_1\tau+u_2\tau^2)(1+u_1^*\tau+u_2^*\tau^2)\nonumber\\& +O(\tau^3)\nonumber\\
&=
1+(u_1+u_1^*)\tau + (u_2+u_2^*+u_1u_1^*)\tau^2 + O(\tau^3).
\label{eqapp:multiply}
\end{align}
Since $M$ is Hermitian, $M_{00}\in\mathbb{R}$, so $u_1=-iM_{00}$ and $u_1+u_1^*=0$. Also $u_1u_1^*=(-iM_{00})(+iM_{00})=M_{00}^2$, and
\begin{align}
(M^2)_{00}&=\sum_{\gamma\in S\cup E}M_{0\gamma}M_{\gamma 0}
= M_{00}^2+\sum_{e\in E}M_{0e}M_{e0}\nonumber\\
&= M_{00}^2+\sum_{e\in E}|M_{0e}|^2.
\end{align}
Therefore $u_2+u_2^*=- (M^2)_{00} = -(M_{00}^2+\sum_{e\in E}|M_{0e}|^2)$, and the $\tau^2$
coefficient in Eq.~(\ref{eqapp:multiply}) becomes
\begin{align}
u_2+u_2^*+u_1u_1^*
&=
-\Bigl(M_{00}^2+\sum_{e\in E}|M_{0e}|^2\Bigr)+M_{00}^2\nonumber\\
&=
-\sum_{e\in E}|M_{0e}|^2.
\end{align}
Hence
\begin{equation}
a(\tau)=1-\tau^2\sum_{e\in E}|M_{0e}|^2+O(\tau^3).
\label{eqapp:atau}
\end{equation}

\paragraph{Step 3: expansion of $b(\tau)=\sum_{e}|U_{0e}(\tau)|^2 n_e$.}
From Eq.~(\ref{eqapp:offdiag}),
\begin{align}
U_{0e}(\tau)&=-iM_{0e}\tau + O(\tau^2);
\\\
U_{0e}^*(\tau)&=+iM_{e0}\tau + O(\tau^2),
\end{align}
so
\begin{align}
|U_{0e}(\tau)|^2 &= U_{0e}(\tau)U_{0e}^*(\tau)
=
(-iM_{0e}\tau)(+iM_{e0}\tau)+O(\tau^3)\nonumber\\
&=
\tau^2 M_{0e}M_{e0}+O(\tau^3)
=
\tau^2|M_{0e}|^2+O(\tau^3).
\label{eqapp:utau}
\end{align}
Substituting Eq.~(\ref{eqapp:utau}) into Eq.~(\ref{eqapp:btau}) yields
\begin{equation}
b(\tau)=\tau^2\sum_{e\in E}|M_{0e}|^2 n_e + O(\tau^3).
\label{eqapp:btau_new}
\end{equation}

\paragraph{Step 4: recovering the short-time RI update formula for $\Delta P$.}
Using the affine map $P_{n+1}=a(\tau)P_n+b(\tau)$, the one-cycle change is
\begin{equation}
\Delta P \equiv P_{n+1}-P_n = \bigl[a(\tau)-1\bigr]P_n + b(\tau).
\label{eqapp:deltap}
\end{equation}
Inserting Eqs.~(\ref{eqapp:atau}) and (\ref{eqapp:btau_new}) into Eq.~(\ref{eqapp:deltap}) gives
\begin{align}
\Delta P
&=
\Bigl[-\tau^2\sum_{e\in E}|M_{0e}|^2+O(\tau^3)\Bigr]P_n \nonumber\\
&+
\Bigl[\tau^2\sum_{e\in E}|M_{0e}|^2 n_e+O(\tau^3)\Bigr]\nonumber\\
&=
-\tau^2\sum_{e\in E}|M_{0e}|^2\bigl(P_n-n_e\bigr)+O(\tau^3),
\label{eqapp:deltaP_new}
\end{align}
which is exactly Eq.~(\ref{eq:DeltaP_general}) obtained earlier from the BCH/double-commutator expansion.

\paragraph{Consistency check (unitarity).}
Row unitarity of $U(\tau)$ implies $|U_{00}(\tau)|^2+\sum_{e\in E}|U_{0e}(\tau)|^2=1$. Using Eqs.~(\ref{eqapp:atau}) and (\ref{eqapp:utau}), we indeed have
\begin{align}
|U_{00}(\tau)|^2+\sum_{e}|U_{0e}(\tau)|^2
&=
\Bigl[1-\tau^2\sum_{e}|M_{0e}|^2\Bigr]\nonumber\\&+\tau^2\sum_{e}|M_{0e}|^2+O(\tau^3)\nonumber\\&=1+O(\tau^3),
\end{align}
as required.

\paragraph{Short-time decay out of an initially occupied fermionic level.}
If $P_0\simeq 1$ and $n_e$ are fixed bath occupations, then for small $n\tau$ one has $P_n\simeq 1$,
so Eq.~(\ref{eqapp:deltaP_new}) gives $\Delta P \simeq -\tau^2\sum_e|M_{0e}|^2(1-n_e)$, reproducing Eq.~(\ref{eq:Gamma_RI_fermion}) of the main text. (For the bosonic analogue, one must distinguish net occupation relaxation from the decay of a tagged excitation; the main-text replacement $1-n_e\to 1+n_e$ corresponds to stimulated emission into occupied modes.)

\section{Generator for the EC protocol}
\label{app:EC_generator}

Here we sketch the derivation of Eqs.~\eqref{eq:EC_block_SS_main}--\eqref{eq:EC_block_SE_main} for the continuous-reset generator in the EC protocol. The starting point is the first-order expansion of the stroboscopic map~\eqref{eq:stroboscopic_map},
\begin{equation}
    V[n+1] = D(\tau)\,V[n] + C(\tau),
\end{equation}
where $V$ is the vector of SPDM components labeled by pairs $(\alpha_i,\beta_i)\in K$ (kept entries), and the stroboscopic map
coefficients are
\begin{align}
    D_{ij}(\tau) &= U^*_{\alpha_i\alpha_j}(\tau)\,U_{\beta_i\beta_j}(\tau);
    \label{eq:app_D_def}\\[2pt]
    C_i(\tau) &= \sum_{(\alpha',\beta')\in R}
    U^*_{\alpha_i\alpha'}(\tau)\,U_{\beta_i\beta'}(\tau)\,
    \rho^{(0)}_{\alpha'\beta'}.
    \label{eq:app_C_def}
\end{align}
Here $U(\tau) = e^{-iM\tau}$ is the single-particle propagator and $\rho^{(0)}$ is the reset SPDM in the block $R$.

We now expand for small $\tau$. Using
\begin{equation}
    U_{\mu\nu}(\tau) = \delta_{\mu\nu} - i\tau M_{\mu\nu} + O(\tau^2),
\end{equation}
we obtain
\begin{align}
    U^*_{\mu\nu}(\tau)
    &= \delta_{\mu\nu} + i\tau M_{\mu\nu}^* + O(\tau^2).
\end{align}
Since $M$ is Hermitian, $M_{\mu\nu}^* = M_{\nu\mu}$.

\paragraph{Expansion of $D_{ij}(\tau)$.}
Substituting into Eq.~\eqref{eq:app_D_def} gives
\begin{align}
    D_{ij}(\tau)
    &= \big(\delta_{\alpha_i\alpha_j} + i\tau M_{\alpha_j\alpha_i}\big)
       \big(\delta_{\beta_i\beta_j} - i\tau M_{\beta_i\beta_j}\big)
       + O(\tau^2)\notag\\
    &= \delta_{\alpha_i\alpha_j}\delta_{\beta_i\beta_j}\notag\\
    &+ i\tau\big(M_{\alpha_j\alpha_i}\delta_{\beta_i\beta_j}
                - \delta_{\alpha_i\alpha_j}M_{\beta_i\beta_j}\big)
    + O(\tau^2).
\end{align}
Identifying $\delta_{\alpha_i\alpha_j}\delta_{\beta_i\beta_j}=\delta_{ij}$, we
have
\begin{equation}
    D_{ij}(\tau)
    = \delta_{ij}
    + i\tau\big(M_{\alpha_j\alpha_i}\delta_{\beta_i\beta_j}
                - \delta_{\alpha_i\alpha_j}M_{\beta_i\beta_j}\big)
    + O(\tau^2).
\end{equation}
Thus the generator matrix $\mathsf{D}$ appearing in
\begin{equation}
    V[n+1] = \big(I + \tau\,\mathsf{D} + O(\tau^2)\big)V[n] + \tau\,\mathsf{C} + O(\tau^2)
\end{equation}
is
\begin{equation}
    \mathsf{D}_{ij}
    = \left.\frac{d}{d\tau}D_{ij}(\tau)\right|_{\tau=0}
    = i\big(M_{\alpha_j\alpha_i}\delta_{\beta_i\beta_j}
            - \delta_{\alpha_i\alpha_j}M_{\beta_i\beta_j}\big).
\end{equation}

\paragraph{Expansion of $C_i(\tau)$.}
From Eq.~\eqref{eq:app_C_def},
\begin{align}
    C_i(\tau)
    &= \sum_{(\alpha',\beta')\in R}
       \big(\delta_{\alpha_i\alpha'} + i\tau M_{\alpha'\alpha_i}\big)
       \big(\delta_{\beta_i\beta'} - i\tau M_{\beta_i\beta'}\big)
       \rho^{(0)}_{\alpha'\beta'}\notag\\
       &+ O(\tau^2)\notag\\
    &= \sum_{(\alpha',\beta')\in R}
       \delta_{\alpha_i\alpha'}\delta_{\beta_i\beta'}\rho^{(0)}_{\alpha'\beta'}
       \notag\\
    &\quad + i\tau\sum_{(\alpha',\beta')\in R}
       \rho^{(0)}_{\alpha'\beta'}
       \big(M_{\alpha'\alpha_i}\delta_{\beta_i\beta'}
           - \delta_{\alpha_i\alpha'}M_{\beta'\beta_i}\big)\notag\\
       &+ O(\tau^2).
\end{align}
Since $(\alpha_i,\beta_i)\in K$ and $R$ is disjoint from $K$, the zeroth-order
term vanishes:
\begin{equation}
    \sum_{(\alpha',\beta')\in R}
       \delta_{\alpha_i\alpha'}\delta_{\beta_i\beta'}\rho^{(0)}_{\alpha'\beta'}
    = 0.
\end{equation}
Therefore
\begin{align}
    C_i(\tau)
    &= i\tau\sum_{(\alpha',\beta')\in R}
       \rho^{(0)}_{\alpha'\beta'}
       \big(M_{\alpha'\alpha_i}\delta_{\beta_i\beta'}
           - \delta_{\alpha_i\alpha'}M_{\beta'\beta_i}\big)\notag\\
      &+ O(\tau^2),
\end{align}
and the generator vector $\mathsf{C}$ is
\begin{align}
    \mathsf{C}_i
    &= \left.\frac{d}{d\tau} C_i(\tau)\right|_{\tau=0}\notag\\
    &= i\sum_{(\alpha',\beta')\in R}
       \rho^{(0)}_{\alpha'\beta'}
       \big(M_{\alpha'\alpha_i}\delta_{\beta_i\beta'}
           - \delta_{\alpha_i\alpha'}M_{\beta'\beta_i}\big).
\end{align}

Finally, rewriting the vector equation for $V$ back in terms of SPDM components $\rho_{\alpha\beta}$ and collecting entries into the system, environment, and coherence blocks reproduces the block equations Eqs.~\eqref{eq:EC_block_SS_main}-\eqref{eq:EC_block_SE_main} in Sec.~\ref{sec:EC}.

\section{Single-level dynamics in the EC protocol}
\label{app:EC_single_level}

We derive the integro-differential equation \eqref{eq:EC_P_memory_cont_main} and the Markovian rate
\eqref{eq:EC_rate_equation_main} for the evolving-correlation (EC) protocol in the single-level case.

\subsection{Block equations and specialization to a single level}

As discussed in Sec.~\ref{sec:EC}, the continuous EC dynamics is obtained by replacing $\rho_{EE}(t)$ with the fixed reference block $\rho_{EE}^{(0)}$ in the unitary equation $\dot{\rho}=-i[M,\rho]$. Writing
\begin{equation}
  \rho(t) =
  \begin{pmatrix}
    \rho_{SS}(t) & \rho_{SE}(t)\\
    \rho_{ES}(t) & \rho_{EE}(t)
  \end{pmatrix};
  \qquad
  M =
  \begin{pmatrix}
    M_{SS} & M_{SE}\\
    M_{ES} & M_{EE}
  \end{pmatrix},
\end{equation}
this gives the block equations
\begin{align}
  \dot{\rho}_{SS}
  &= -i[M_{SS},\rho_{SS}]
     -i\big(M_{SE}\rho_{ES}-\rho_{SE}M_{ES}\big);
  \label{eq:EC_block_SS_app}\\
  \dot{\rho}_{SE}
  &= -i\big(
        M_{SS}\rho_{SE}+M_{SE}\rho_{EE}^{(0)} \notag\\
        &-\rho_{SS}M_{SE}-\rho_{SE}M_{EE}
      \big),
  \label{eq:EC_block_SE_app}
\end{align}
with $\rho_{ES}=\rho_{SE}^\dagger$. We now restrict to a single system level, labelled by $0\in S$, and an environment that has been diagonalized in the eigenbasis of $M_{EE}$:
\begin{equation}
  M_{EE}\ket{k} = \omega_k\ket{k},\qquad k\in\mathcal{I}.
\end{equation}
The system--environment couplings are $g_k\equiv M_{0k}$, and the reference environment state is taken to be diagonal in this basis with
\begin{equation}
  (\rho_{EE}^{(0)})_{kk'} = n_k\delta_{kk'}.
\end{equation}

We denote the occupation of the system level by
\begin{equation}
  P(t) \equiv \rho_{00}(t),
\end{equation}
and define the coherences
\begin{equation}
  s_k(t) \equiv \rho_{0k}(t),\qquad
  \rho_{k0}(t)=s_k^*(t).
\end{equation}
Since $M_{SS}=\omega_0$ is a $1\times1$ scalar, the commutator $[M_{SS},\rho_{SS}]$ vanishes identically. Evaluating the $(0,0)$ and $(0,k)$ matrix elements of Eqs.~\eqref{eq:EC_block_SS_app}–\eqref{eq:EC_block_SE_app}
then gives
\begin{align}
  \dot{P}(t)
  &= -i\big(M_{0k}\rho_{k0}(t)-\rho_{0k}(t)M_{k0}\big)\notag\\
  &= -i\sum_k\big(g_k s_k^*(t)-s_k(t)g_k^*\big);
  \label{eq:EC_P_dot_app}\\
  \dot{s}_k(t)
  &= -i\Big(
        \omega_0 s_k(t)
        + g_k n_k
        - P(t)\,g_k
        - s_k(t)\omega_k
      \Big)\notag\\
  &= -i\big[\Delta_k s_k(t) + g_k\big(n_k-P(t)\big)\big],
  \label{eq:EC_sk_dot_app}
\end{align}
where $\Delta_k\equiv\omega_0-\omega_k$. Equation~\eqref{eq:EC_sk_dot_app} is a
linear inhomogeneous equation for $s_k(t)$.

\subsection{Integral equation for the survival probability}

Assuming $s_k(0)=0$ for all $k$, the solution of
Eq.~\eqref{eq:EC_sk_dot_app} reads
\begin{equation}
  s_k(t)
  = -i g_k \int_0^t dt'\,e^{-i\Delta_k(t-t')}\big(n_k-P(t')\big),
\end{equation}
and hence
\begin{equation}
  s_k^*(t)
  =  i g_k^* \int_0^t dt'\,e^{+i\Delta_k(t-t')}\big(n_k-P(t')\big).
\end{equation}
Substituting these expressions into Eq.~\eqref{eq:EC_P_dot_app} yields
\begin{align}
  g_k s_k^*(t) - s_k(t)g_k^*
  &= i|g_k|^2 \int_0^t dt'\,e^{+i\Delta_k(t-t')}\big(n_k-P(t')\big)\notag\\
  &\quad
     + i|g_k|^2 \int_0^t dt'\,e^{-i\Delta_k(t-t')}\big(n_k-P(t')\big)\notag\\
  &= i2|g_k|^2 \int_0^t dt'\,
         \cos\big[\Delta_k(t-t')\big]\notag\\ &\times\big(n_k-P(t')\big).
\end{align}
Inserting this into Eq.~\eqref{eq:EC_P_dot_app} we obtain
\begin{align}
  \dot{P}(t)
  &= -i\sum_k\big(g_k s_k^*(t)-s_k(t)g_k^*\big)\notag\\
  &= -2\sum_k |g_k|^2 \int_0^t dt'\,
        \cos\big[\Delta_k(t-t')\big]\big(P(t')-n_k\big).
  \label{eq:EC_P_memory_app}
\end{align}
This is the desired integro-differential equation for the survival probability in the EC protocol.

Introducing the spectral density
\begin{equation}
  J(\omega) = \sum_k |g_k|^2\delta(\omega-\omega_k),
\end{equation}
and the occupation function $n(\omega)$ defined by $n(\omega_k)=n_k$, Eq.~\eqref{eq:EC_P_memory_app} can be written in continuum form as
\begin{align}
  \dot{P}(t)
  = -2\int d\omega\,J(\omega)\int_0^t dt'\,
      &\cos\big[(\omega_0-\omega)(t-t')\big]\notag\\
      &\times \big(P(t')-n(\omega)\big),
  \label{eq:EC_P_memory_cont_app}
\end{align}
which coincides with Eq.~\eqref{eq:EC_P_memory_cont_main} in the main text.

\subsection{Markovian limit and effective decay rate}

We now make the standard weak-coupling and smooth-spectrum approximations. The memory kernel in Eq.~\eqref{eq:EC_P_memory_cont_app} decays on a time scale set by the inverse bandwidth of $J(\omega)$, whereas the survival probability $P(t)$ varies slowly on the relaxation time scale. We therefore approximate $P(t')\simeq P(t)$ inside the integral and extend the upper integration limit
in $\int_0^t d(t-t')$ to infinity:
\begin{equation}
  \int_0^\infty d\tau\,\cos\big[(\omega_0-\omega)\tau\big]
  = \pi\delta(\omega_0-\omega),
\end{equation}
where principal-value contributions produce only a Lamb shift of $\omega_0$, which we neglect. Equation~\eqref{eq:EC_P_memory_cont_app} then reduces to
\begin{align}
  \dot{P}(t)
  &\simeq -2\big(P(t)-n(\omega_0)\big)
      \int d\omega\,J(\omega) \notag\\
      &\times\int_0^\infty d\tau\,\cos\big[(\omega_0-\omega)\tau\big]\notag\\
  &= -2\big(P(t)-n(\omega_0)\big)
      \int d\omega\,J(\omega)\,\pi\delta(\omega_0-\omega)\notag\\
  &= -2\pi J(\omega_0)\big(P(t)-n(\omega_0)\big).
\end{align}
Thus we obtain the Markovian rate equation
\begin{equation}
  \dot{P}(t)
  = -\Gamma_{\mathrm{EC}}\big(P(t)-n(\omega_0)\big),
  \qquad
  \Gamma_{\mathrm{EC}} = 2\pi J(\omega_0),
\end{equation}
whose solution is
\begin{equation}
  P(t)
  = n(\omega_0) + \big(P(0)-n(\omega_0)\big)e^{-\Gamma_{\mathrm{EC}}t}.
\end{equation}
The effective decay rate $\Gamma_{\mathrm{EC}}$ depends only on the local bath spectral density at the system frequency and is independent of the reset interval $\tau$ in the continuous EC limit. Consequently, the EC protocol does not exhibit strict Zeno freezing, in contrast to the RI protocol where $\Gamma_{\mathrm{RI}}(\tau)\propto\tau$ as $\tau\to 0$.

\bibliography{references_doi_audited}

\end{document}